\documentclass{emulateapj}

\pdfoutput=1 	
\usepackage{epsf,amsfonts,amsmath,amssymb}
\usepackage{float}
\usepackage[pagebackref=false]{hyperref}
\usepackage[all]{hypcap} 
\usepackage{cleveref}
\usepackage{graphicx}
\usepackage{dcolumn}
\usepackage{bm}
\usepackage{multirow}

\graphicspath{{C:/Users/bnhan3764/MyResearch/publications/cosmic_twilight_polarimeter_ctp/induced_foregnd_polarization/Figures/}}
\newcommand{\myreferences}{C:/Users/bnhan3764/MyResearch/publications/cosmic_twilight_polarimeter_ctp/induced_foregnd_polarization/Reference}

\newcommand\T{\rule{0pt}{2.6ex}}   
\newcommand\B{\rule[-1.2ex]{0pt}{0pt}}

\begin{document}
\title{A polarimetric approach for constraining the dynamic foreground
  spectrum \\for cosmological global 21-cm measurements}

\author{Bang D. Nhan\altaffilmark{1,2,5}, Richard
  F. Bradley\altaffilmark{2,3,4}, and Jack O. Burns\altaffilmark{1}}

\altaffiltext{1}{Center for Astrophysics and Space
  Astronomy (CASA), Department of Astrophysical and Planetary
  Sciences,\\ University of Colorado, Boulder, CO 80309, USA}
\altaffiltext{2}{National Radio Astronomy Observatory (NRAO), Central
  Development Laboratory (CDL), Charlottesville, VA 22903, USA}
\altaffiltext{3}{Department of Astronomy, University of Virginia,
  Charlottesville, VA 22903, USA} 
\altaffiltext{4}{Department of Electrical and Computer Engineering,
  University of Virginia, Charlottesville, VA 22903, USA}
\altaffiltext{5}{NRAO Grote Reber Doctoral Research Fellow at CDL,
  \texttt{bang.nhan@colorado.edu}}

\date{\today}

\begin{abstract}
  The cosmological global (sky-averaged) 21-cm signal is a powerful
  tool to probe the evolution of the intergalactic medium (IGM) in
  high-redshift Universe ($z \leq 6$). One of the biggest
  observational challenges is to remove the foreground spectrum which
  is at least four orders of magnitude brighter than the cosmological
  21-cm emission. Conventional global 21-cm experiments rely on the
  spectral smoothness of the foreground synchrotron emission to
  separate it from the unique 21-cm spectral structures in a single
  total-power spectrum. However, frequency-dependent instrumental and
  observational effects are known to corrupt such smoothness and
  complicates the foreground subtraction. We introduce a polarimetric
  approach to measure the projection-induced polarization of the
  anisotropic foreground onto a stationary dual-polarized antenna. Due
  to Earth rotation, when pointing the antenna at a celestial pole,
  the revolving foreground will modulate this polarization with a
  unique frequency-dependent sinusoidal signature as a function of
  time. In our simulations, by harmonic decomposing this dynamic
  polarization, our technique produces two separate spectra in
  parallel from the same observation: (i) a total sky power consisting
  both the foreground and the 21-cm background, (ii) a
  model-independent measurement of the foreground spectrum at a
  harmonic consistent to twice the sky rotation rate. In the absence
  of any instrumental effects, by scaling and subtracting the latter
  from the former, we recover the injected global 21-cm model within
  the assumed uncertainty. We further discuss several limiting factors
  and potential remedies for future implementation.

\end{abstract}


\maketitle

\section{Introduction}
\label{sec:introduction}
A theoretical framework has established three primary transition
phases in the early Universe, namely the Cosmic Dark Ages (a period
between recombination at redshift $z \sim 1,100$ and the formation of
the first stars $z \gtrsim 30$), the Cosmic Dawn (when the first stars
ignited, $30 \gtrsim z \gtrsim 15$), and the Epoch of Reionization
(EoR, when the first stars and galaxies reionized the neutral hydrogen
HI into singly ionized hydrogen HII between $15 \gtrsim z \gtrsim 6$)
\citep{furlanetto2006global, pritchard2010constraining,
  pritchard201221, liu2013global, mirocha2013interpreting}. While the
current theoretical models of these phases are compelling, their
parameter spaces are still unconstrained by any direct
observations. Detailed physics on how the first stars, black holes,
and galaxies form from the primordial neutral hydrogen reservoir in
the intergalactic medium (IGM) is still not well understood.

The highly redshifted HI 21-cm hyperfine line (rest frequency $\nu_0
\sim$ 1,420 MHz) provides a promising observational probe to study the
high-$z$ Universe where few or none of the luminous objects are
available \citep{furlanetto2006cosmology}. This hyperfine transition
happens at the $1S$ ground state of the hydrogen atom when the
magnetic dipole moments of the proton and electron flip from parallel
(triplet state) to antiparallel (singlet state). Theory suggests that
the redshifted 21-cm brightness temperature fluctuations $\delta
T_{b,21cm}(z)$ measured against the cosmic microwave background (CMB)
can be parametrized as \citep{furlanetto2006global},
\begin{equation}
  \begin{aligned}
    \delta T_{b,21cm}(z) =&\ 27(1-x_i)(1+\delta)
    \left(\frac{\Omega_{b,0}h^2}{0.023}\right)\\
    & \times \left(\frac{0.15}{\Omega_{m,0}h^2}\frac{1+z}{10}\right)^{1/2}
    \left(\frac{T_S-T_{\gamma}}{T_S}\right) {\rm mK},
    \end{aligned}
\label{eq:21cm_deltb}
\end{equation}
where $x_i$ is the ionized fraction of HI, $\delta$ is the fractional
overdensity, $h$ is the Hubble parameter today in unit of 100 km
s$^{-1}$ Mpc$^{-1}$, $\Omega_{b,0}$ and $\Omega_{m,0}$ are the
fractional contributions of baryons and matter to the critical energy
density, $T_S$ is the HI's spin temperature and $T_{\gamma} =
T_{\gamma,0}(1+z)$ is the CMB temperature at $z$ with present day
value $T_{\gamma,0} = 2.725$ K. The 21-cm transition is observed in
emission when $T_S > T_{\gamma}$ and absorption otherwise. As the
Universe evolves, the HI content is being depleted and the thermal
environment is altered. Hence, being able to measure how the
redshifted 21-cm signal strength evolves over time at the low
frequency regime ($\nu \lesssim 200$ MHz), we can constrain the
neutral fraction and thermal history in the IGM of the early Universe.

Most of the ground-based observational efforts in the past decades are
using large interferometer arrays to measure the power spectrum of
spatial fluctuations of the 21-cm brightness temperature at the end of
the EoR. Such experiments include the Murchison Widefield Array
\citep[MWA, ][]{tingay2013murchison, bowman2013science}, the Donald
C. Backer Precision Array for Probing the Epoch of Reionization
\citep[PAPER, ][]{parsons2010precision}, the Low Frequency Array
\citep[LOFAR, ][]{van2013lofar}, the Giant Metrewave Radio Telescope
\citep[GMRT, ][]{paciga2013simulation}, the Square Kilometer Array
\citep[SKA, ][]{mellema2013reionization}, and the Hydrogen Epoch of
Reionization Array \citep[HERA, ][]{pober2014what}.

\begin{figure}
\includegraphics[scale=.59]{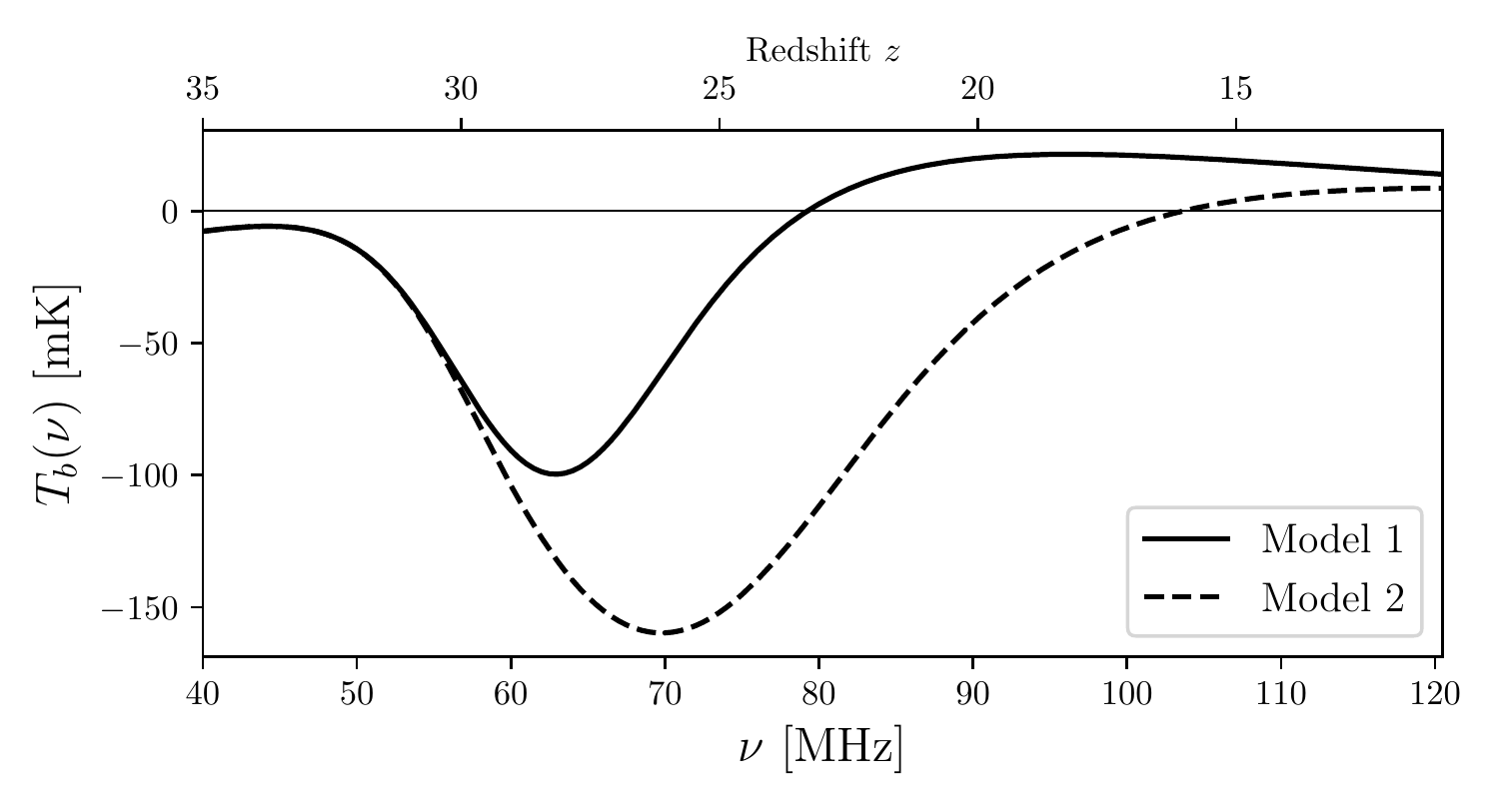} 
\caption{Two examples of simulated global 21-cm cosmology signal
  relative to the CMB. These models are generated using the
  Accelerated Reionization Era Simulations (ARES) code with fiducial
  model (solid curve, Model 1) and a factor of 10 less in the rate of
  X-ray heating (dashed curve, Model 2). Spectral structures of the
  21-cm signal help to constrain the thermal history of the early
  universe between $z$ of 35-11, which corresponds to observed
  frequencies of 40-120 MHz.}\label{fig:model_signal_paper}
\end{figure}

An emerging alternative approach is to utilize a single dipole antenna
or a compact array consisting of a small number of antenna elements to
capture the 21-cm signal evolution over a large range of frequencies
($\sim 40 \le \nu \le 200$ MHz) for the aforementioned transition
phases \citep{pritchard2010constraining, liu2013global}. The observed
signal is spatially averaged by the broad antenna beams and is
commonly known as the global (or sky-averaged) 21-cm signal. More
specifically, theoretical models suggest that there are unique
emission and absorption features imprinted on the global 21-cm
spectrum which signify the time and physical conditions for each of
the transition phases. The peak absolute amplitude of this broadband
signal is predicted to range from a few tens to hundreds of
millikelvin, depending on particular star formation model parameters
chosen. For examples, \autoref{fig:model_signal_paper} illustrates how
the brightness temperature $T_b(\nu)$ varies as a function of observed
frequency for two global 21-cm models, with a fiducial model (Model 1)
and a factor of 10 less in the rate of X-ray heating from the
universe's first galaxies (Model 2), both generated by the Accelerated
Reionization Era Simulations \citep[ARES, ][]{mirocha2014decoding}. In
this particular simulation, decreasing only the X-ray heating rate
produces an absorption trough at higher frequency. Details on how
different astrophysical parameters may alter the shape of the global
21-cm spectrum, using the similar simulation, are discussed in other
studies \citep[e.g.,][]{mirocha2013interpreting,
  mirocha2015interpreting, mirocha2017global}.

At present, some of the existing ground-based global 21-cm experiments
are the Experiment to Detect the Global Epoch of Reionization
Signature \citep[EDGES, ][]{bowman2010lower}, the Large-Aperture
Experiment to Detect the Global the Dark Ages \citep[LEDA,
][]{greenhill2015constraining}, the Shaped Antenna Measurement of the
Background Radio Spectrum \citep[SARAS, ][]{patra2013saras}, the Sonda
Cosmologica de las Islas para la Deteccion de Hidrogeno Neutro
\citep[SCI-HI, ][]{voytek2014probing}, and the Broadband Instrument
for Global Hydrogen Reionization Signal \citep[BIGHORN,
][]{sokolowski2015bighorns}. Space-based mission concepts such as the
Dark Age Radio Explorer \citep[DARE, ][]{burns2012probing} are also
being proposed to carry out high precision measurements in orbit above
the lunar farside for a more pristine radio environment.

Challenges arising in both observational approaches include, but not
limited to, the removal of foreground emission originated from
Galactic and extragalactic radio sources, ionospheric distortion,
antenna beam pattern effects, instrumental systematic constraints, and
terrestrial radio-frequency interference (RFI) from the weak
cosmological 21-cm signal. Among these, the foreground emission is the
strongest contaminant since it has an averaged brightness temperature
at least four orders of magnitude stronger than the predicted 21-cm
signal \citep{shaver1999can}.

At low frequencies, the foreground emission is dominated by
non-thermal synchrotron radiation originated from high energy
cosmic-ray (CR) electrons interacting with magnetic field in the
Galaxy and other extragalactic sources
\citep{kogut2012synchrotron}. Motivated by the theory of diffusive
shock acceleration (or Fermi acceleration) which predicts a power-law
energy distribution for the charged particles moving in magnetized gas
clouds, the observed foreground spectrum is generally parametrized as
a spectrally smooth power law with a running spectral index
\citep{shaver1999can, tegmark2000foregrounds,
  kogut2012synchrotron}. One of a more commonly adopted
parametrization is a polynomial in log-log space
\citep{pritchard2010constraining, bowman2010lower, harker2012mcmc} as,
\begin{equation}
\log \hat{T}_{fg}(\nu) = \sum_{n=0}^{m>0}c_n\left(\log \nu\right)^n
\label{eq:log_poly}
\end{equation}
where $\hat{T}_{fg}(\nu)$ is the estimated foreground spectrum with
polynomial coefficient $c_n$ of order $n$. In principle, the estimated
foreground can be subtracted from the measured spectrum to reveal the
global 21-cm spectrum. Instrumental systematics and other
observational variations complicate the global signal extraction.

This foreground removal approach is statistically justified to extract
the weak 21-cm spectrum that has distinct small-scaled frequency
structure in contrast to the smoothly varying foreground
\citep{petrovic2011systematic}. However, this approach is limited by
having to infer that the cosmological signal and foreground emission
simultaneously in a single total-power spectrum. Unless the foreground
spectrum is measured to a precision within few millikelvins,
subtracting such a bright foreground directly from the
observed-frequency space will introduce errors that are equivalent to
the global 21-cm signal level \citep{petrovic2011systematic,
  harker2015selection}. In addition, free-free emission, recombination
lines, and instrumental biases can also distort the spectral
smoothness of the foreground spectrum. In fact, recent studies
\citep{bernardi2015foreground, mozdzen2016limits} find that nonlinear
chromatic effects from antenna beam structures can distort the
smoothness of the foreground spectrum and further complicate the
removal process.

It is desirable to have a direct means to separate and constrain the
foreground spectrum from the 21-cm signal in the same observation with
minimal dependence on any foreground model. Efforts in isolating the
foreground from the background signal observationally have been
explored in previous studies. For example, the PAPER and HERA
experiments adopt the delay filtering technique in an attempt to
separate the foreground in delay space instead of frequency space
\citep{parsons2012per}. For the global signal, a singular value
decomposition (SVD) analysis is used to construct suitable spectral
basis to capture the time variable component of the measured dynamic
spectra to constrain the foreground spectrum
\citep{vedantham2013chromatic}. Meanwhile, a principle component
analysis (PCA) is adopted to determine contaminated spectral modes in
the smooth foreground spectrum based on its spatial fluctuations
\citep{switzer2014erasing}.

In this study, we propose an observational technique that also
utilizes the dynamic nature of the foreground to separate it from the
static background through polarimetry. Projection of the anisotropic
foreground source distribution induces a net polarization on a
stationary dual-polarized antenna. The apparent relative motion
between the Earth and the foreground modulates a unique periodic
waveform to the polarization measurements that helps to constrain the
foreground spectrum independently of the homogeneous background 21-cm
signal. Similar modulation techniques by rotating the instrument to
separate the polarization properties of instrument relative to the
astrophysical sources have been shown to be effective in CMB
measurements \citep[e.g.,][]{johnson2007maxipol,
  kusaka2014modulation}.  Through simulations, we demonstrate how the
foreground spectrum is isolated using the projection-induced
polarization and helps to extract the global 21-cm signal.

In \autoref{sec:general_theory}, the general theory and rationales are
presented on how the projection-induced foreground polarization can
help to measure the foreground spectrum. This is followed by the
mathematical formulation and an analytical example in
\autoref{sec:foreground_separation}. Simulation results using a
realistic sky map are presented in \autoref{sec:haslam_sim}. In
\autoref{sec:implementation_practicality}, we discuss some of
practical challenges in implementing the polarimetric approach. We
conclude in \autoref{sec:conclusion} with a summary of key findings
and future plans for implementing a polarimeter prototype.

\section{General theory and rationales}
\label{sec:general_theory}
Spatially, a majority of the diffuse foreground emission is
concentrated on the Galactic plane, with extragalactic sources
sparsely distributed above and below it. In the global 21-cm
experiment, with the large antenna beam width, the 21-cm fluctuations
with small angular scale \citep[$<2^{\circ}$,
][]{bittner2011measuring} are not resolvable comparing to the
foreground anisotropy with larger angular scale. Hence, the observed
signal is a linear combination of a relatively isotropic 21-cm
background and the foreground emission.

Since conventional global 21-cm experiments have the antennas pointed
toward the zenith, the foreground spectrum varies as a function of
Local Sidereal Time (LST). In combination with ionospheric distortion
\citep{vedantham2013chromatic, datta2016effects, sokolowski2015impact,
  rogers2015radiometric} and RFI \citep{offringa2013brightness}, this
temporal variations can complicate the accuracy of constraining the
foreground spectrum over the course of the observation. To mitigate
this, the sky measurement is averaged over a small time window,
typically several sidereal hours per day to obtain a mean-valued
total-power spectrum.

By recognizing the strong spatial variations in the foreground in
contrast to the isotropic background, we propose a polarimetric
approach that can measure the foreground spectrum without relying on
any presumed sky model. For a given sky region, projection of
anisotropic foreground onto the antenna plane will produce a net
composite polarization. Meanwhile, the isotropic cosmological 21-cm
background is not polarized due to symmetry. Noting that this
projection-induced polarization is distinct from the well-known
intrinsic polarization of the Galactic synchrotron emission. The Earth
rotation will modulate the magnitude of the induced polarization and
couple a strong periodic structure to the polarization measurement as
a function time. The proposed technique is to exploit this dynamic
characteristic in the induced polarization as a direct means to
distinguish the foreground from the background.

Our approach consists of pointing a stationary dual-polarized antenna,
a pair of crossed dipoles above a finite ground plane, at a fixed
reference point in the sky about which the same sky region is observed
continuously. For a ground-based experiment, this unique sky pointing
can either be the North Celestial Pole (NCP) for the northern
hemisphere or the South Celestial Pole (SCP) for the southern,
  assuming the effects of precession of the equinoxes is negligible.
Without loss of generality, as the Earth rotates, the field of view
(FOV) for the sky about the NCP is constant throughout the entire
observation and the foreground regions appear to revolve about these
two points.

The measured composite projection-induced polarization can be
parametrized in terms of a net Stokes vector $\bold{S}_{net}(\nu,t)$
as a function of time across the frequency range of interest, where
$\bold{S}_{net}(\nu,t) = \left\{I_{net}^{\nu}(t), Q_{net}^{\nu}(t),
U_{net}^{\nu}(t), V_{net}^{\nu}(t)\right\}$. The total sky intensity
at each frequency, including both the background and foreground, is
represented by the Stokes $I_{net}^{\nu}(t)$. The Stokes
$I_{net}^{\nu}(t)$ is a linearly combination of the unpolarized
($I_u^{\nu}$) and polarized ($I_p^{\nu}$) portion of the sky signal,
hence $I_p(t) \leq I_{net}(t)$.

\begin{figure}
  \includegraphics[scale=0.6]{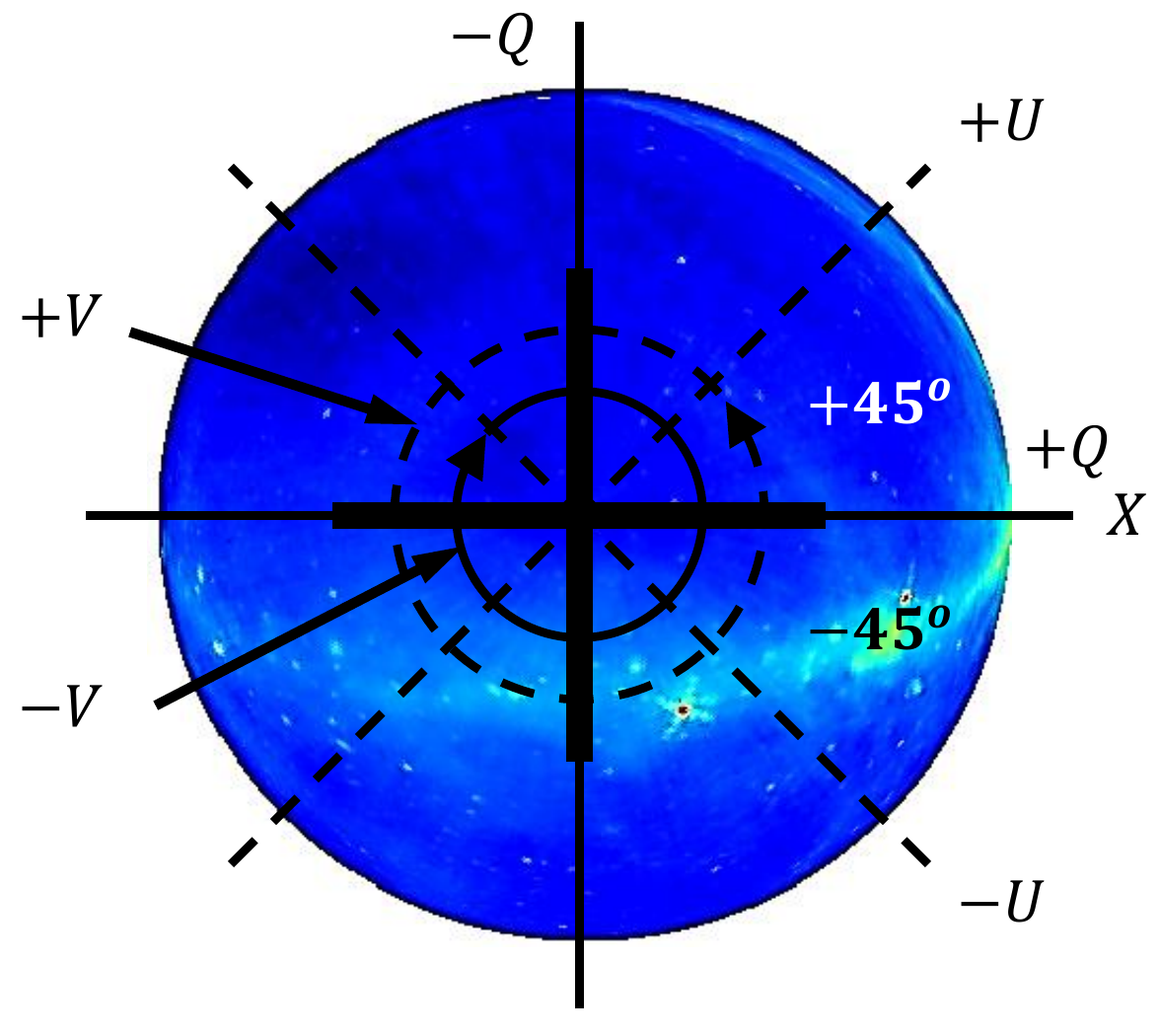} 
  \caption{Aligning the crossed dipoles to the celestial pole, the
    projection-induced polarization from the foreground anisotropy can
    be characterized by a net composite Stokes vector in Poncair\'e
    space. The corresponding Stokes parameters $Q, U,$ and $V$ measure
    different orientations of the polarization: $\pm Q$ for
    $0^{\circ}$ and $90^{\circ}$, $\pm U$ for $\pm 45^{\circ}$, $+V$
    for left-handed, and $-V$ for right-handed.
  }\label{fig:stokes_vector_illustration}
\end{figure}

By convention, the Stokes
$\left\{Q_{net}^{\nu}(t),U_{net}^{\nu}(t),V_{net}^{\nu}(t)\right\}$ can
be mapped in a 3-D Cartesian coordinate to form a sphere (Poincar\'e
sphere) of radius $R = I_p = \sqrt{Q_{net}^2 + U_{net}^2 +
  V_{net}^2}$. The Stokes $Q_{net}(t)$ measures linear polarizations
at $0^{\circ}$ and $90^{\circ}$, whereas the Stokes $U_{net}(t)$
measures linear polarizations at $\pm 45^{\circ}$ relative to the
antenna's orientation as shown in
\autoref{fig:stokes_vector_illustration}. The Stokes $V_{net}(t)$ can
measure any existing circular polarization with positive and negative
values for left-handed and right-handed orientation, respectively.

Because the same net foreground polarization is observed twice by the
same dipole after the sky revolves by 180-degree apart, the linear Stokes
$Q_{net}(t)$ and $U_{net}(t)$ are sinusoidal functions with an angular
frequency of twice the sky revolution rate relative to the fixed
antenna. This twice-angular frequency in the Stokes $Q_{net}(t)$
and $U_{net}(t)$ is the dynamic feature that helps to distinguish the
foreground component from the static 21-cm background. By applying
harmonic analysis such as Fourier transformation to estimate the power
spectral density (PSD) of these two Stokes parameters for each of the
observed frequency channels, a second harmonic associated with twice
the angular frequency component can be determined. The amplitude of this
second harmonic can be compiled together, channel by channel, to
construct a replica of the foreground spectrum. We refer this
constructed spectrum as the second-harmonic Stokes spectrum hereafter.

Meanwhile, the constant total intensity $I_{net}^{\nu}(t)$ contains
both the background 21-cm signal and foreground emission, as in the
case of the conventional total-power measurement approach, is the
zero-frequency component in the PSD. In a similar manner as above, the
total power Stokes spectrum can be constructed. As a result, the
zeroth harmonic Stokes spectrum contains both the background and
foreground signal, whereas the second-harmonic Stokes spectrum only
captures information about the foreground. In principle, scaling and
subtracting the second-harmonic Stokes spectrum from the total-power
spectrum can recover the embedded 21-cm signal. Mathematical details
and simulations are presented in the following sections.

\section{Foreground spectrum constraints with projection-induced polarization}
\label{sec:foreground_separation}
\subsection{Stokes formalism}
\label{sec:stoke_para_formalism}
Consider a coordinate system fixed to the sky and a pair of
crossed-dipole antennas with their boresight aligned along the
$z-$axis at $\theta = 0^{\circ}$ and the horizon is at $\theta =
90^{\circ}$ on the $xy-$plane for the antenna's coordinate system. We
further assume that the sky coordinate is not observable below the
horizon at $\theta \leq 90^{\circ}$, although the ionosphere can
extend the visible sky below the horizon due to refraction in the low
frequencies \citep[e.g., ][]{vedantham2013chromatic}. From a given
direction $(\theta,\phi)$ in the sky at a given instant, the broadband
incoming unpolarized radio signal can be represented as
quasi-monochromatic electric field with polar and azimuthal components
$\bold{E}_{in}(\theta,\phi,\nu) =
\{E_{\theta}(\theta,\phi,\nu)\bold{\hat{\theta}} +
E_{\phi}(\theta,\phi,\nu)\bold{\hat{\phi}}\}$. The incoming $E$-field
is projected and received by the crossed dipoles, which have an $X$
and $Y$ oriented $E$-fields that are orthogonal to each other. The
observed field, $\bold{E}_{out}(\theta,\phi,\nu)$ in Cartesian
coordinates for such a system, is a product of the incoming field with
a Jones matrix \citep{trippe2014polarization},
\begin{equation}
  \begin{aligned}
    \bold{E}_{out}(\theta,\phi,\nu) &= E_{X}(\theta,\phi,\nu)\bold{\hat{x}} + E_{Y}(\theta,\phi,\nu)\bold{\hat{y}}\\
                  &= \left[\bold{J}_{ant}^{X}(\theta,\phi,\nu) + 
                  \bold{J}_{ant}^{Y}(\theta,\phi,\nu)\right] \bold{E}_{in}(\theta,\phi,\nu) \\
                  &= \bold{J}_{ant}(\theta,\phi,\nu)\bold{E}_{in}(\theta,\phi,\nu)
  \end{aligned}
\label{eq:projection_efield_out}
\end{equation}
where the superscripts on the Jones matrices indicate the $X$ and $Y$
dipoles. In a realistic system, the Jones matrix also describes other
instrumental systematics as well as external influences on the
electric field such as ionospheric effects \citep{kraus1986radio,
  heiles2001heuristic, trippe2014polarization}. Here, we only consider
the projection of the incoming $E$-field onto the dipole antennas,
which is described by $\bold{J}_{ant}$.

The antenna Jones matrix $\bold{J}_{ant}$ is the composite far-field
patterns of both crossed dipoles at each frequency, which is
represented as,
\begin{equation}
  \begin{aligned}
    \bold{J}_{ant}(\theta,\phi,\nu) &= 
    \begin{bmatrix}
      J_{\theta}^{X} & J_{\phi}^{X}\\
      J_{\theta}^{Y} & J_{\phi}^{Y}
    \end{bmatrix}_{(\theta,\phi,\nu)}\\
    &= 
    \begin{bmatrix}
      |E_{\theta}^{X}|e^{j\Phi_{\theta}^{X}} & |E_{\phi}^{X}|e^{j\Phi_{\phi}^{X}} \\
      |E_{\theta}^{Y}|e^{j\Phi_{\theta}^{Y}} & |E_{\phi}^{Y}|e^{j\Phi_{\phi}^{Y}}
    \end{bmatrix}_{(\theta,\phi,\nu)}
  \end{aligned}
  \label{eq:jones_matrix}
\end{equation}
where the complex far-field components for the $X$ and $Y$ oriented
dipoles are given in terms of their magnitude $|E|$ and phase $\Phi$
for each direction $(\theta,\phi)$.

Components of the Stokes vector at a given direction can be calculated
as linear combinations of the observed $E$-field's autocorrelation and
cross-correlation,
\begin{equation}
  \begin{aligned}
    I_{out}^{\nu}(\theta,\phi) &= \langle E_XE_X^*\rangle + \langle E_YE_Y^*\rangle \\
    Q_{out}^{\nu}(\theta,\phi) &= \langle E_XE_X^*\rangle - \langle E_YE_Y^*\rangle\\
    U_{out}^{\nu}(\theta,\phi) &= \langle E_XE_Y^*\rangle + \langle E_X^*E_Y\rangle\\
    V_{out}^{\nu}(\theta,\phi) &= j\left(\langle E_XE_Y^*\rangle - \langle E_X^*E_Y\rangle\right)
  \end{aligned}
\label{eq:stoke_parameters_individual}
\end{equation}
where the bracket represents a time average of the product between the
observed $E$-field components and their complex conjugates, with $j =
\sqrt{-1}$. The spatial and frequency dependence on the $E$-field are
suppressed for the ease of reading.

Assuming the $E$-field from all directions add incoherently, the net
composite field measured at the crossed dipoles are sums of squares
instead of the square of sums of individual electric field. Hence, the
Stokes vector representing the projection-induced polarization of the
foreground at a given instant is a vector summation of the individual
Stokes vectors from each direction on the Poincar\'e sphere,
\begin{equation}
  \begin{aligned}
    I_{net}^{\nu} &= \frac{1}{\Omega_{XY}}\sum^{2\pi}_{\phi=0}\sum^{\pi/2}_{\theta=0}I_{out}^{\nu}(\theta,\phi)\Delta\Omega\\
    Q_{net}^{\nu} &= \frac{1}{\Omega_{XY}}\sum^{2\pi}_{\phi=0}\sum^{\pi/2}_{\theta=0}Q_{out}^{\nu}(\theta,\phi)\Delta\Omega\\
    U_{net}^{\nu} &= \frac{1}{\Omega_{XY}}\sum^{2\pi}_{\phi=0}\sum^{\pi/2}_{\theta=0}U_{out}^{\nu}(\theta,\phi)\Delta\Omega\\
    V_{net}^{\nu} &= \frac{1}{\Omega_{XY}}\sum^{2\pi}_{\phi=0}\sum^{\pi/2}_{\theta=0}V_{out}^{\nu}(\theta,\phi)\Delta\Omega\\
  \end{aligned}
\label{eq:stoke_parameters_net}
\end{equation}
where $\Omega_{XY} =
\sum^{2\pi}_{\phi=0}\sum^{\pi/2}_{\theta=0}F(\theta,\phi,\nu)\Delta\Omega$
is the beam normalization factor of the averaged antenna radiation
pattern $F(\theta,\phi,\nu) = (F^X + F^Y)/2$, where
$F^{X,Y}(\theta,\phi,\nu) = |E_{\theta}^{X,Y}(\theta,\phi,\nu)|^2 +
|E_{\phi}^{X,Y}(\theta,\phi,\nu)|^2$ are the radiation patterns of
antenna $X$ and $Y$, with the differential solid angle element
$\Delta\Omega = \sin\theta\Delta\theta\Delta\phi$.

\subsection{Dynamic characteristic of the projection-induced polarization}
\label{sec:dynamic_polarization}
Since the sky revolves about the celestial pole periodically every
sidereal day, this dynamic characteristic is also carried over to the
resulting projection-induced polarization. To illustrate this, we
simplify the projection by aligning the crossed dipoles' boresight
with the NCP at $\theta=0^{\circ}$. This is analogous to placing the
antenna at the Geographic North Pole (GNP, at latitude $\phi_{\oplus}
= 90^{\circ}$), such that the same sky revolves about the celestial
pole continuously at an angular frequency $\omega_{sky}$, where
$\omega_{sky}/2\pi \ll \nu$.

Because the coordinate system we adopted that is fixed to the sky, the
$(\theta,\phi)$ coincide with the declination DEC and right ascension
RA in the equatorial system, respectively. As the sky revolves
relative to the fixed antennas on the ground, the net induced
polarization reaches its maximal or minimal values when it is parallel
or orthogonal to one of the dipoles. For the same dipole, the net
induced polarization is detected twice per revolution, or twice
diurnal. This gives rise to the cyclic Stokes parameters
$Q_{net}^{\nu}(t)$ and $U_{net}^{\nu}(t)$ with an angular frequency
equaling twice the sky revolving rate, i.e., $2\omega_{sky}$.

In such a coordinate system, the observed $E$-field is produced as if
having the antenna rotated relative to the fixed sky. By reciprocity,
we can model the antenna response coupled to the incoming $E$-field
from direction $(\theta,\phi)$ as one emitted by an infinitesimal
horizontal crossed dipoles placed at the origin of the coordinate
system. For the dipole along $x$-axis, the complex farfield measured
at distance $r$ has components,
\begin{equation}
 \begin{aligned}
   E_{\theta}^X(\theta,\phi,\nu) &\simeq -j\frac{\omega\mu I_0 l e^{-j\bold{k}\cdot\bold{r}}}{4\pi r}\cos\theta\cos\phi \\
   E_{\phi}^X(\theta,\phi,\nu)  &\simeq +j\frac{\omega\mu I_0 l e^{-j\bold{k}\cdot\bold{r}}}{4\pi r}\sin\phi     
 \end{aligned}
\label{eq:farfield_X}
\end{equation}
where $\omega = 2\pi\nu$, $|\bold{k}| = 2\pi/\lambda$ is the wave
number of the observed wavelength, $\mu$ is the permeability, and
$I_0$ is the current excited across the infinitesimal dipole with
length $l$ by the source \citep{balanis2016antenna}. Similarly, for
the $y$-oriented dipole, components of the corresponding farfield are,
\begin{equation}
 \begin{aligned}
   E_{\theta}^Y(\theta,\phi,\nu) &\simeq -j\frac{\omega\mu I_0 l e^{-j\bold{k}\cdot\bold{r}}}{4\pi r}\cos\phi\\
   E_{\phi}^Y(\theta,\phi,\nu)   &\simeq -j\frac{\omega\mu I_0 l e^{-j\bold{k}\cdot\bold{r}}}{4\pi r}\cos\theta\sin\phi 
 \end{aligned}
\label{eq:farfield_Y}
\end{equation}

As the sky revolves about the boresight of the antenna over some time
$t$, a sky emission from direction $(\theta_0,\phi_0)$, that is
originally coupled to the antenna response at $(\theta_0,\phi_0)$, is
now received to the antenna response $(\theta',\phi') =
(\theta_0,\phi_0 + \omega_{sky}t)$. By substituting the fields from
Equation \eqref{eq:farfield_Y} into Equation
\eqref{eq:projection_efield_out} to
\eqref{eq:stoke_parameters_individual}, the resulting Stokes
parameters corresponding to $(\theta,\phi)$ is a function of time and
are written as,
\begin{equation}
  \begin{aligned}
    I_{out}^{\nu}(\theta,\phi,t) = &g_0^2\left(E_{\theta}^2\cos^2{\theta} + E_{\phi}^2\right)\\
    Q_{out}^{\nu}(\theta,\phi,t) = &g_0^2\left(E_{\theta}^2\cos^2{\theta} - E_{\phi}^2\right)\cos(2\omega_{sky} t)-\\
                                  &2g_0^2E_{\theta}E_{\phi}\cos{\theta}\sin(2\omega_{sky} t)\\
    U_{out}^{\nu}(\theta,\phi,t) = &g_0^2\left(E_{\theta}^2\cos^2{\theta} - E_{\phi}^2\right)\sin(2\omega_{sky} t)+\\
                                  &2g_0^2E_{\theta}E_{\phi}\cos{\theta}\cos(2\omega_{sky} t)\\
    V_{out}^{\nu}(\theta,\phi,t) = &0
  \end{aligned}
\label{eq:stoke_parameters_ptsrc}
\end{equation}
where $g_0 = \omega\mu I_0 l/4\pi r$, noting that the incoming
$E$-field components are time independent. The net induced
polarization, which is estimated by the vector sum of all Stokes
vectors from all direction above the horizon as Equation
\eqref{eq:stoke_parameters_net}, subsequently consists of this
twice-diurnal component as a function of time. This is the dynamic
feature that we exploit to measure the foreground separately from the
static background.

It is trivial that the net polarization can become relatively small,
if not zero, when the overall foreground anisotropy decreases. For
example, if the incoming $E$-field components are identical for all
direction, the vector sum of the Stokes vectors reduces to zero. This
is similar to the symmetry argument for the isotropic background
signal. As a result, the magnitude of twice diurnal component can be
altered. However, this is very unlikely in a real observation except
for some extreme cases, such as horizon obstruction at low latitude
discussed in \autoref{sec:effect_horizon}. A more comprehensive
analysis is elaborated through simulations with a realistic foreground
map in \autoref{sec:haslam_sim} and discussions in
\autoref{sec:implementation_practicality}.

\subsection{Harmonic analysis and Stokes spectra}
\label{sec:harmonic_analysis}
For a continuous signal $g(t)$, its PSD can be estimated by the
Fourier decomposition in the frequency domain. For discrete signal
data with length of $N$ and sampling interval $\Delta t$, one of the
optimized PSD estimates \citep{heinzel2002spectrum} is,
\begin{equation}
  S_g^{\nu}(f) = \frac{(\Delta t)^2}{s_1^2}\left|\sum_{t=1}^{N}w(t)g(t)e^{-j2\pi ft}\right|^2
    \label{eq:stoke_psd}
\end{equation}
where the window function $w(t)$ prevents spectral leakage between
frequency channels in the PSD, and the normalization factor $s_1 =
\sum_{t=0}^{N}w(t)$.  In this paper, we adopt the commonly used
Blackman-Harris window function for the PSD calculations. At each
observed frequency, a Stokes PSD can be estimated by replacing $g(t)$
with each of the Stokes parameters. Here we explicitly denote the
dynamic frequency as $f$ to distinguish it from the observed frequency
$\nu$. In the context of the sky rotation, $f = \omega_{sky}/2\pi$.

A harmonic of order $n$ will appear in the PSD if $g(t)$ consists of a
frequency component at $f$, where $n = 1/(fN\Delta t)$. Hence, the
available power associated to the components of twice the angular
frequency in Stokes $Q_{net}(t)$ and $U_{net}(t)$ are estimated as
$S_Q^{\nu}(n=2)$ and $S_U^{\nu}(n=2)$. Meanwhile, in case of the
constant total intensity from $I_{net}^{\nu}(t)$, the power is located
at the zeroth harmonic $S_{I}^{\nu}(n=0)$.

Since the $\nu$-dependence of the Stokes parameters is not affected by
the Fourier transformation, the spectral dependence of the sky
measurement is carried over to the Stokes PSD. The measured
$S_Q^{\nu}(n=2)$ and $S_U^{\nu}(n=2)$ originated from the dynamic
component of the foreground polarization are model-independent,
high-fidelity, and scalable replicas of the foreground spectrum, which
do not contain any contribution from the isotropic 21-cm
background. The static background combined with the foreground is
measured at the zero-frequency component of the total intensity
spectrum $S_{I}^{\nu}(n=0)$.

\subsection{Foreground subtraction with induced Stokes spectra}
\label{sec:foreground_subtraction}
The sky-averaged antenna temperature $T_{ant}(\nu)$ for a given
antenna radiation pattern $F(\theta,\phi,\nu)$ is defined as, 
\begin{equation}
  T_{ant}(\nu) = \frac{
    \int_{0}^{2\pi}\int_{0}^{\pi/2} T_{sky}(\nu,\theta,\phi)F(\theta,\phi,\nu)\sin\theta{\rm
      d}\theta{\rm d}\phi}{\int_{0}^{2\pi}\int_{0}^{\pi/2} F(\theta,\phi,\nu)\sin\theta{\rm
      d}\theta{\rm d}\phi}
  \label{eq:T_ant_def}
\end{equation}
where $T_{sky}(\nu,\theta,\phi)$ is the brightness temperature
distribution of the total sky \citep{kraus1986radio,
  wilson2009tools}. In an ideal case, the sky temperature is simply a
linear combination of the foreground emission $T_{fg}(\nu)$ and the
background 21-cm signal $\delta T_{b,21cm}(\nu)$. The total-power
spectrum at zero-frequency $S_{I}^{\nu}(n=0)$ is equivalent to
$T_{ant}(\nu)$ and can be parametrized as,
\begin{equation}
  S_{I,0}^{\nu} \equiv S_{I}^{\nu}(n=0) = A_1 k_B\left[T_{fg}(\nu) + \delta T_{b,21cm}(\nu)\right] + A_0
  \label{eq:stokeI_spec_n0}
\end{equation}
Similarly, the second-harmonic Stokes spectra are,
\begin{equation}
  \begin{aligned}
    S_{Q,2}^{\nu} \equiv S_Q^{\nu}(n=2) &= B_1k_BT_{fg}(\nu) + B_0 \\
    S_{U,2}^{\nu} \equiv S_U^{\nu}(n=2) &= C_1k_BT_{fg}(\nu) + C_0
  \end{aligned}
  \label{eq:stokeQU_spec_n2}
\end{equation}
where the $A$s, $B$s, and $C$s are the scaling factors and offsets. In
practice, these scaling coefficients are corrupted by instrumental
effects and become frequency-dependent. By assuming optimal instrument
calibration, the coefficients can be treated as constants for the
purpose of this paper.

Since all the Stokes spectra are measured simultaneously in the same
observation, by solving for the scaling factors, the foreground
spectrum can be subtracted to extract the background signal, without
the need of high-order polynomial fitting using, for instance,
Equation \eqref{eq:log_poly}. Although either equation in
Equation \eqref{eq:stokeQU_spec_n2} can be used to solve for $T_{fg}$
because both Stokes $Q_{net}(t)$ and $U_{net}(t)$ measure the same
linear polarization but with a phase difference of $90^{\circ}$, using
both measurements at the same time can improve the precision of the
$n=2$ harmonic since we know both Stokes parameters share the same
angular frequency component and a unique relative phase difference of
$90^{\circ}$ apart.

\begin{figure}
\centering 
\includegraphics[scale=.59]{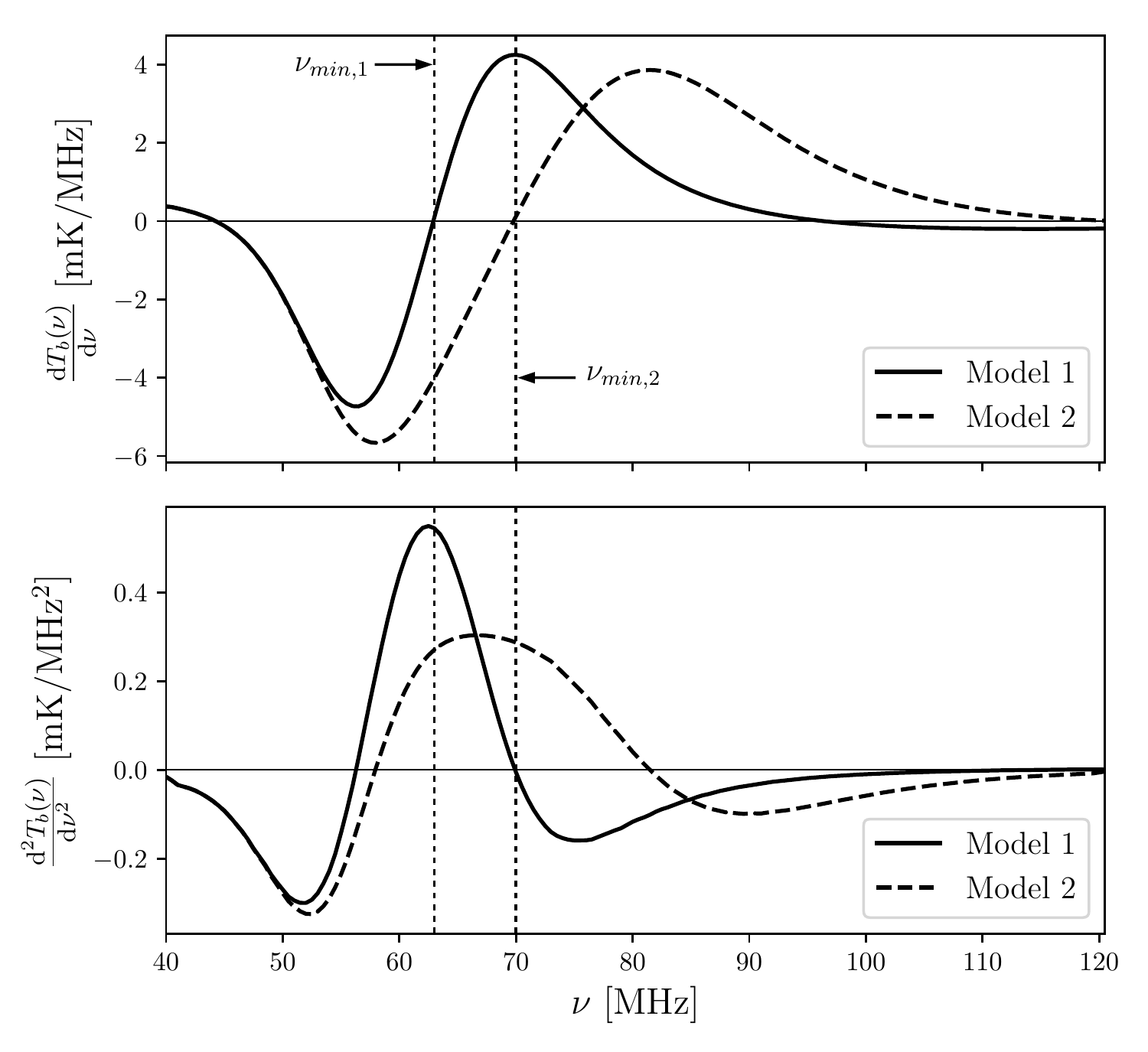}
\caption{\emph{Upper:} Zero crossings of the first derivatives of the
  ARES global 21-cm Model 1 (solid curve) and 2 (dashed curve) from
  \autoref{fig:model_signal_paper} help to determine their global
  minima at $\nu_{min,1}$ and $\nu_{min,2}$ (vertical dashed lines)
  respectively. These global minima help to obtain the coefficients to
  scale the second-harmonic spectrum $S_{Q,2}^{\nu}$ up to the total
  power spectrum $S_{I,0}^{\nu}$ for foreground
  subtraction. \emph{Lower:} Magnitudes of the second derivatives of
  the models help to differentiate potential local minima from the
  global minima required for the scaling
  coefficients.}\label{fig:model_signal_derivatives_paper}
\end{figure}

For simplicity, without loss of generality, by substituting the
$Q$-spectrum into Equation \eqref{eq:stokeI_spec_n0} and solve for the
background signal, so we have,
\begin{equation}
  \delta T_{b,21cm}(\nu) = \frac{1}{k_B}\left(\frac{S_{I,0}^{\nu} - A_0}{A_1} -
  \frac{S_{Q,2}^{\nu}-B_0}{B_1}\right)
  \label{eq:T21cm_extract1}
\end{equation}
though utilizing both Stokes measurement, along with Stokes $V$
  measurement, can be useful in practice. We can obtain a relation
between $A_1$ and $B_1$ by taking the derivative on both sides of
Equation \eqref{eq:T21cm_extract1} as,
\begin{equation}
  \frac{A_1}{B_1} = \biggl[\frac{{\rm d}S_{I,0}^{\nu}}{{\rm d}\nu} -
    A_1k_B\frac{{\rm d}\delta T_{b,21cm}(\nu)}{{\rm d}\nu}\biggr]
  \left(\frac{{\rm d}S_{Q,2}^{\nu}}{{\rm d}\nu}\right)^{-1}
 \label{eq:scaling_coeff0}
\end{equation}
If the global 21-cm spectrum consists of an absorption feature at
$\nu_{min}$ as predicted by the models such as ones shown in
\autoref{fig:model_signal_paper}, there exists a global minimum at
$\nu_{min}$ that can be determined by its first and second derivatives
as shown in \autoref{fig:model_signal_derivatives_paper}. Hence, we
can solve for $A_1/B_1$ based entirely on the derivatives of the
measured Stokes spectra since the derivative of $\delta
T_{b,21cm}(\nu)$ is zero if we evaluate Equation
\eqref{eq:scaling_coeff0} at $\nu_{min}$
\begin{equation}
  \frac{A_1}{B_1} = \frac{{\rm d}S_{I,0}^{\nu}}{{\rm d}\nu} 
  \left(\frac{{\rm d}S_{Q,2}^{\nu}}{{\rm d}\nu}\right)^{-1}\bigg|_{\nu=\nu_{min}}
  \label{eq:scaling_coeff1}
\end{equation} 
In principle, the first derivatives of the emission features at the
higher end of the passband in \autoref{fig:model_signal_paper} also
provide a zero crossing to constrain $A_1/B_1$. However, the emission
feature is more likely to be outside the band since it is predicted to
be broader and less prominent than the absorption one.

By comparing \eqref{eq:scaling_coeff0} and \eqref{eq:scaling_coeff1},
it is apparent that the sources of error in solving for $A_1/B_1$ are
the accuracy in estimating the derivatives of the measured Stokes
spectra as well as determining $\nu_{min}$. In general, there exists
truncation error in numerical differentiation in the order of the
frequency step size $\mathcal{O}(\Delta\nu)$
\citep{hamming2012numerical}. Hence, keeping the spectral resolution
low can improve the accuracy of calculating the derivatives. However,
determining the frequency of the absorption feature in the global
21-cm signal is not straightforward since the background signal is
unknown. Possible ways helping to constrain $\nu_{min}$ and the
sensitivity of this scheme are further discussed in
\autoref{sec:foreground_subtraction_error} and
\autoref{sec:spectra_sensitivity}. It is worth noting that the
foreground removal discussed here only applies when the an absorption
feature exists in the observing band as predicted by theory ($\sim
40<\nu<200$ MHz).

\section{Simulations with a realistic foreground map}
\label{sec:haslam_sim}
\subsection{Simulation description}
\label{sec:sim_description}
In this section, we illustrate how the projection-induced polarization
from a realistic sky map can give rise to the second-harmonic spectrum
and how a model global 21-cm spectrum can be recovered using the
procedure presented in the previous section.

For simplicity, the broad farfield beam patterns of the crossed
dipoles are approximated by circular Gaussian beams. Referencing from
the typical antenna beam sizes from other global 21-cm experiments like
EDGES \citep{mozdzen2016limits}, we adopt a full width at half maximum
(FWHM) of $60^{\circ}$ for the Gaussian beams.

\begin{figure}
\centering 
\includegraphics[scale=.65]{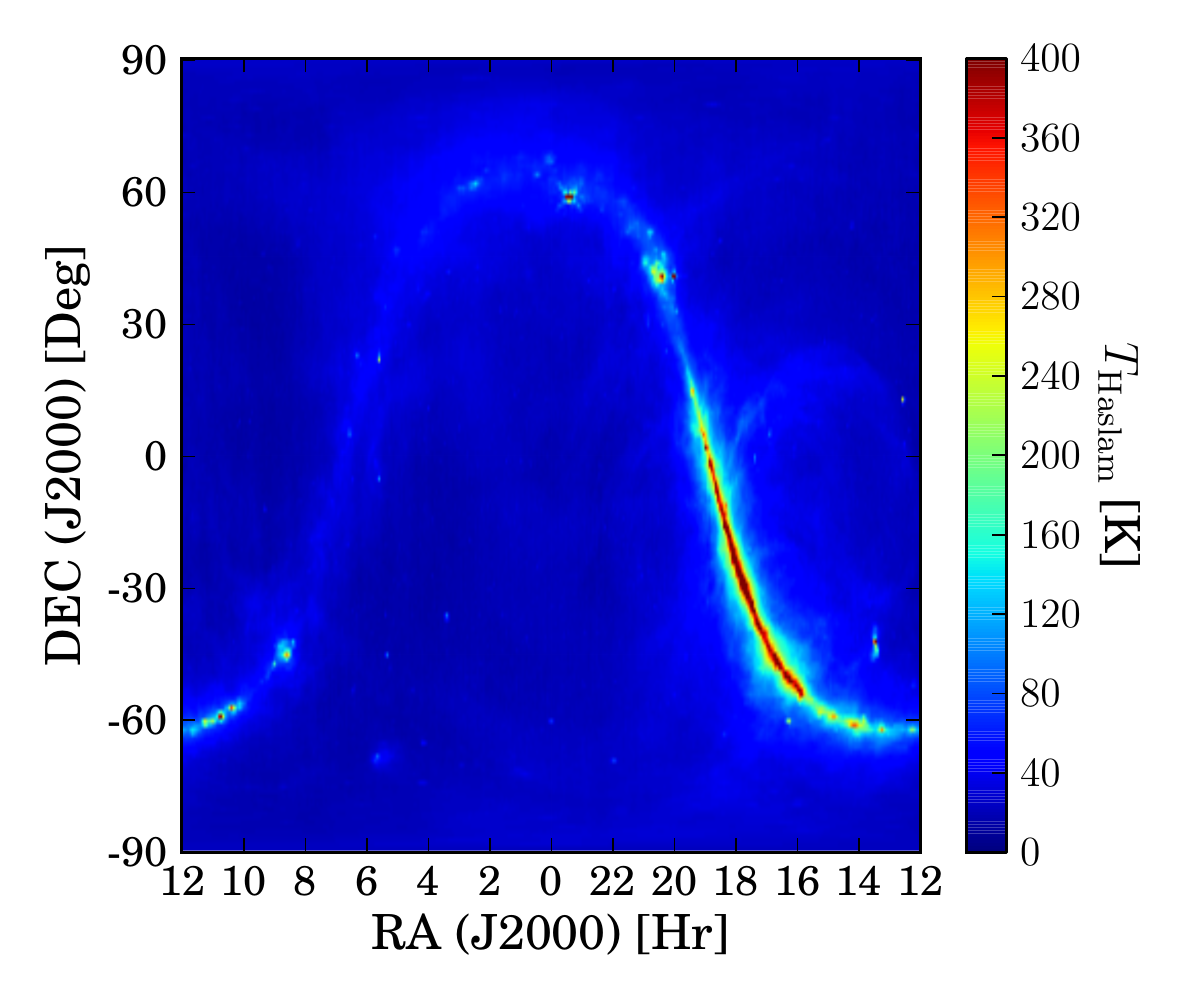} 
\caption{The realistic foreground between 40-120 MHz is extrapolated
  from the Haslam full-sky map at 408 MHz, as shown here, with a spectral index
  $\beta$ of 2.47.}\label{fig:haslam_map_equatorial_grid}
\end{figure}

The Gaussian beams are centered at the NCP of a set of foreground maps
which are extrapolated from the Haslam full-sky survey at 408 MHz, as
shown in \autoref{fig:haslam_map_equatorial_grid}
\citep{haslam1982408}, to 40-120 MHz using a power-law parametrization
with a mean spectral index $\beta = 2.47$ as suggested by previous
observations \citep{bowman2008toward},
\begin{equation}
T_{fg}(\theta,\phi,\nu) = T_{\rm Haslam}(\theta,\phi)\left(\frac{\nu}{408\ {\rm
    MHz}}\right)^{-\beta}
\label{eq:pw_law_fg}
\end{equation}
For each of the ARES global 21-cm model, at each frequency, an
isotropic 21-cm background map of brightness temperature $\delta
T_{b,21cm}(\nu)$ for all directions is linearly combined with the
foreground map $T_{fg}(\nu,\theta,\phi)$. Since the spatial resolution
is not critical in the global 21-cm measurement, it is sufficient to
set the spatial resolution of both the beam patterns and the sky map
to $1^{\circ}\times 1^{\circ}$ in our simulations. We also assume the
Gaussian beams to be frequency independent across the passband, which is
not necessarily true for the real antenna beams.

Although intrinsic linear polarization is known to exist in the
Galactic synchrotron foreground, we assume the foreground itself to be
unpolarized (or randomly polarized) in the simulations since the
sky-averaged intrinsic polarization is expected to be less dominant
than net projection-induced polarization. For completeness, both the
chromatic beam effects and intrinsic polarization are elaborated in
\autoref{sec:implementation_practicality}.

For an unpolarized foreground, half of the averaged total power is
equally received by each of the antenna polarization. According to the
formalism presented in \autoref{sec:stoke_para_formalism}, we estimate
the incoming $E$-field from the foreground as
$\bold{E}_{in}(\theta,\phi,\nu) =
\{E_0(\theta,\phi,\nu)\bold{\hat{\theta}} +
E_0(\theta,\phi,\nu)\bold{\hat{\phi}}\}$ where $E_0(\theta,\phi,\nu) =
\sqrt{k_BT_{fg}(\theta,\phi,\nu)/2}$ with the Boltzmann constant
$k_B$.

\subsection{Simulation results}
\label{sec:sim_results}
\begin{figure}
\centering 
\includegraphics[scale=.475]{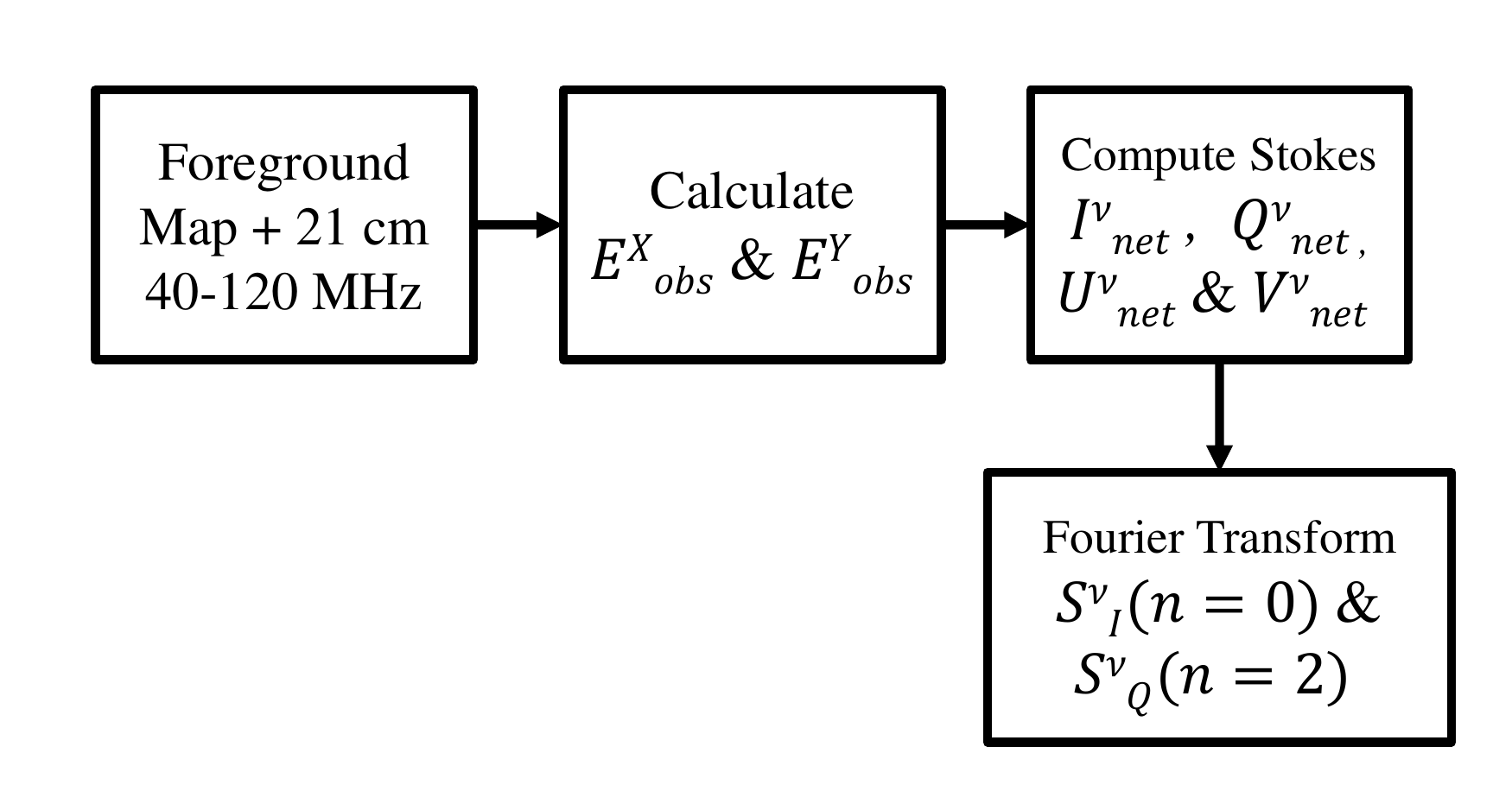}
\caption{Summary for the simulations and analysis procedures that
  allows the foreground spectrum be separated from the isotropic 21-cm
  background by exploiting the dynamic characteristic in the
  projection-induced foreground
  polarization.}\label{fig:ctp_analysis_block_diagram_haslam}
\end{figure}

After the observed $E$-field is calculated as in Equation
\eqref{eq:projection_efield_out}, the net Stokes parameters are
computed as the foreground revolves about the NCP at a rate of one
cycle per 24 sidereal hours. The resulting Stokes parameters are
Fourier transformed to compute the PSD. Zeroth- and second-harmonic
spectra, $S_{I,0}^{\nu}$ and $S_{Q,2}^{\nu}$, are constructed by
assembling the magnitude of the FFT output at $n=0$ and $n=2$ for all
frequencies. \autoref{fig:ctp_analysis_block_diagram_haslam}
summarizes the basic steps to produce the Stokes spectra.

\begin{figure}
\centering 
\includegraphics[scale=.475]{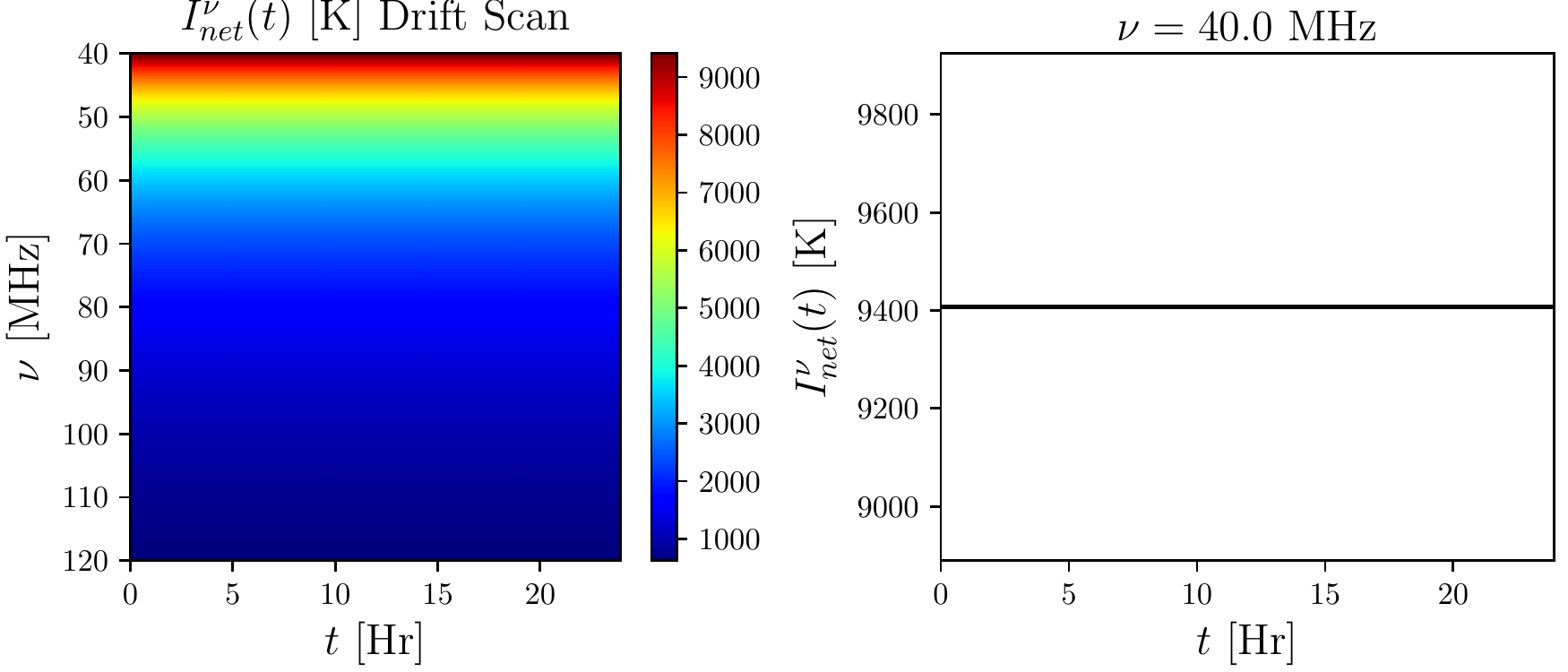} 
\caption{\emph{Left:} Stokes $I_{net}$ drift scan for Haslam
  foreground example with circular Gaussian beams with the time on the
  $x$-axis and the observed frequencies on the $y$-axis. \emph{Right:}
  An example of the Stokes $I_{net}$ which is constant for all time at
  $\nu = 40$ MHz. This total intensity measurement contains both the
  foreground and background
  signal.}\label{fig:stokes_I_drift_scan_gauss_circ_haslam_paper}
\end{figure}

\begin{figure}
\centering 
\includegraphics[scale=.475]{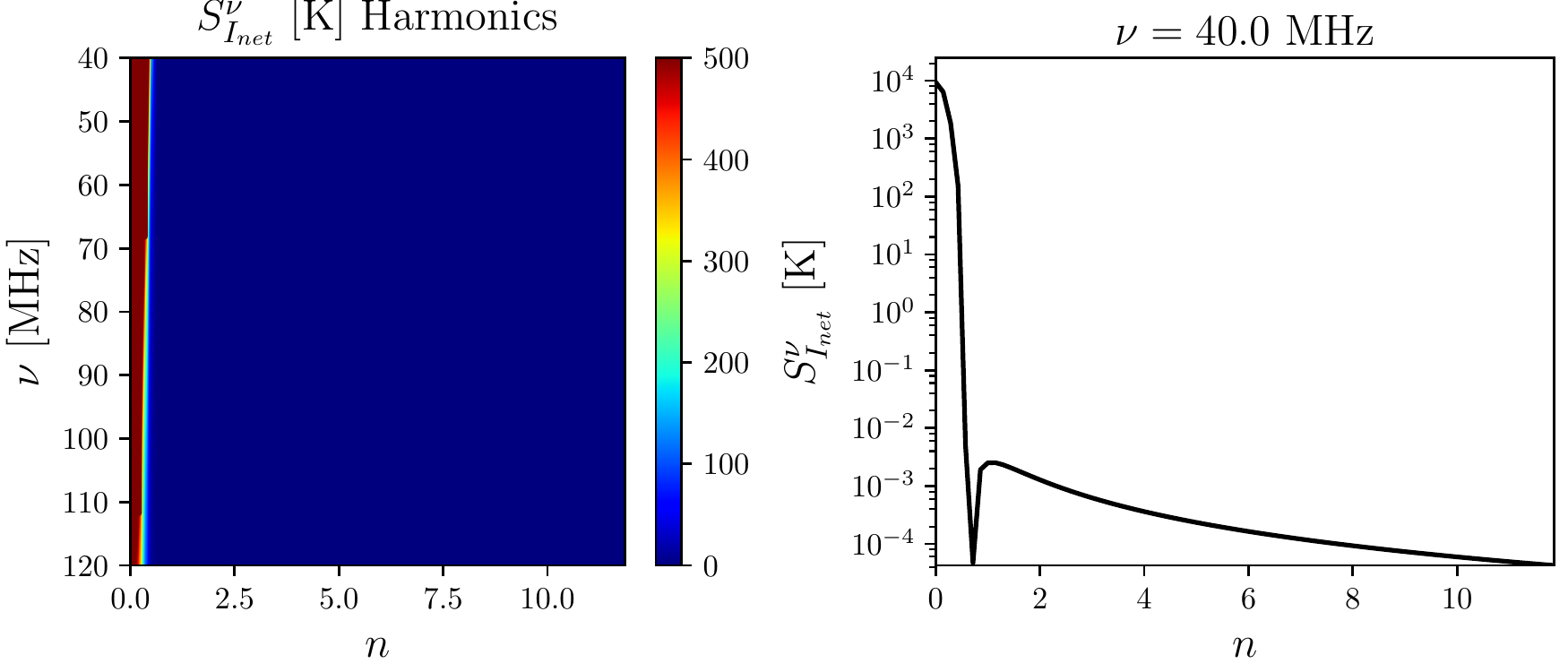} 
\caption{\emph{Left:} The FFT of the Stokes $I_{net}$ are shown in
  terms of harmonic order $n$ as a function of observed
  frequencies. \emph{Right:} The total available power for the constant
  Stokes $I_{net}$ can be find at harmonic $n=0$. The zeroth-harmonic
  Stokes spectrum $S_{I,0}^{\nu}$ is constructed by assembling the
  magnitude at $n=0$ across the 40-120 MHz band.
}\label{fig:stokes_I_harmonic_gauss_circ_haslam_paper}
\end{figure}

\begin{figure}
\centering 
\includegraphics[scale=.475]{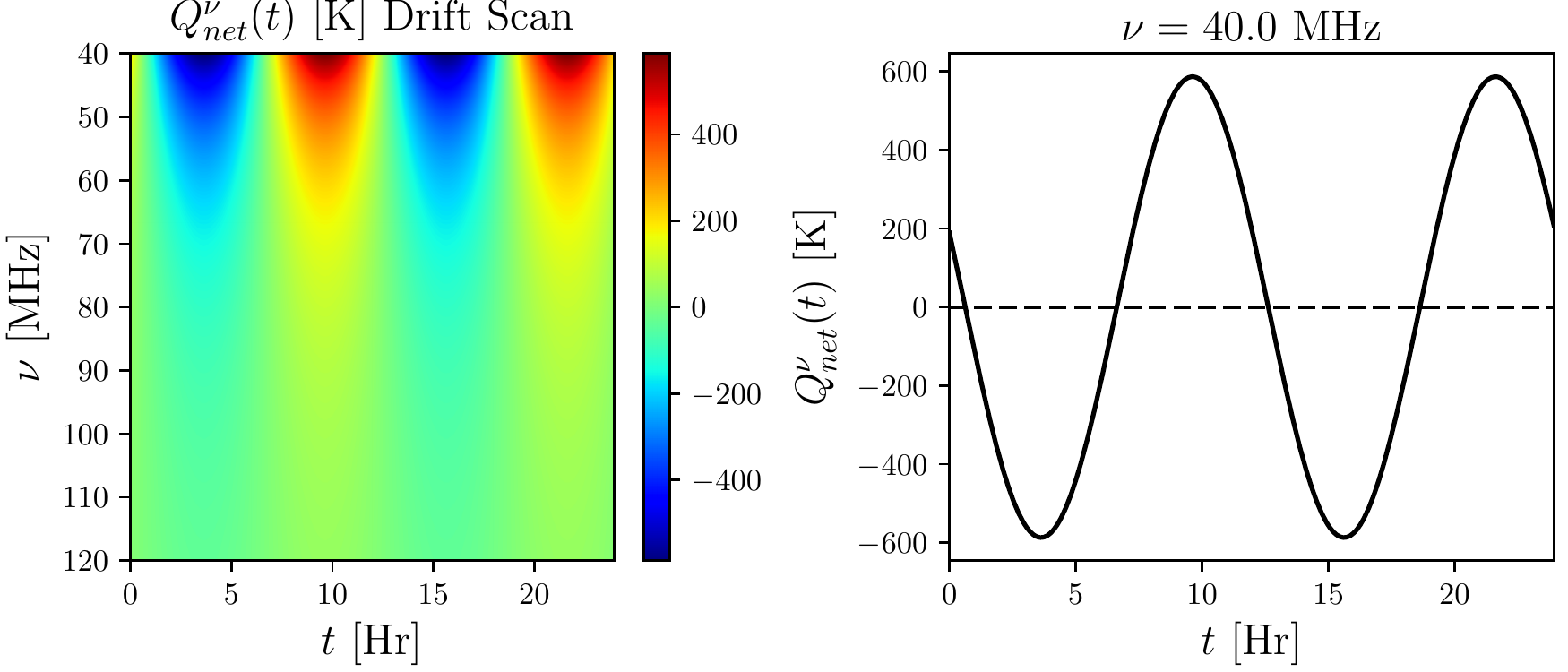} 
\caption{\emph{Left:} Stokes $Q_{net}$ drift scan for Haslam
  foreground example with circular Gaussian beams with the time on the
  $x$-axis and the observed frequencies on the $y$-axis. \emph{Right:}
  The projection-induced polarization is modulated by a sinusoidal
  waveform with angular frequency of twice the sky rotation rate at
  each observed frequency, as shown here for $\nu = 40$ MHz. This
  dynamic characteristic is unique to the foreground and can be used
  to separate it from the static background so the spectral structures
  of the foreground spectrum can be constrained without assuming any
  sky model.  }\label{fig:stokes_Q_drift_scan_gauss_circ_haslam_paper}
\end{figure}

\begin{figure}
\centering 
\includegraphics[scale=.475]{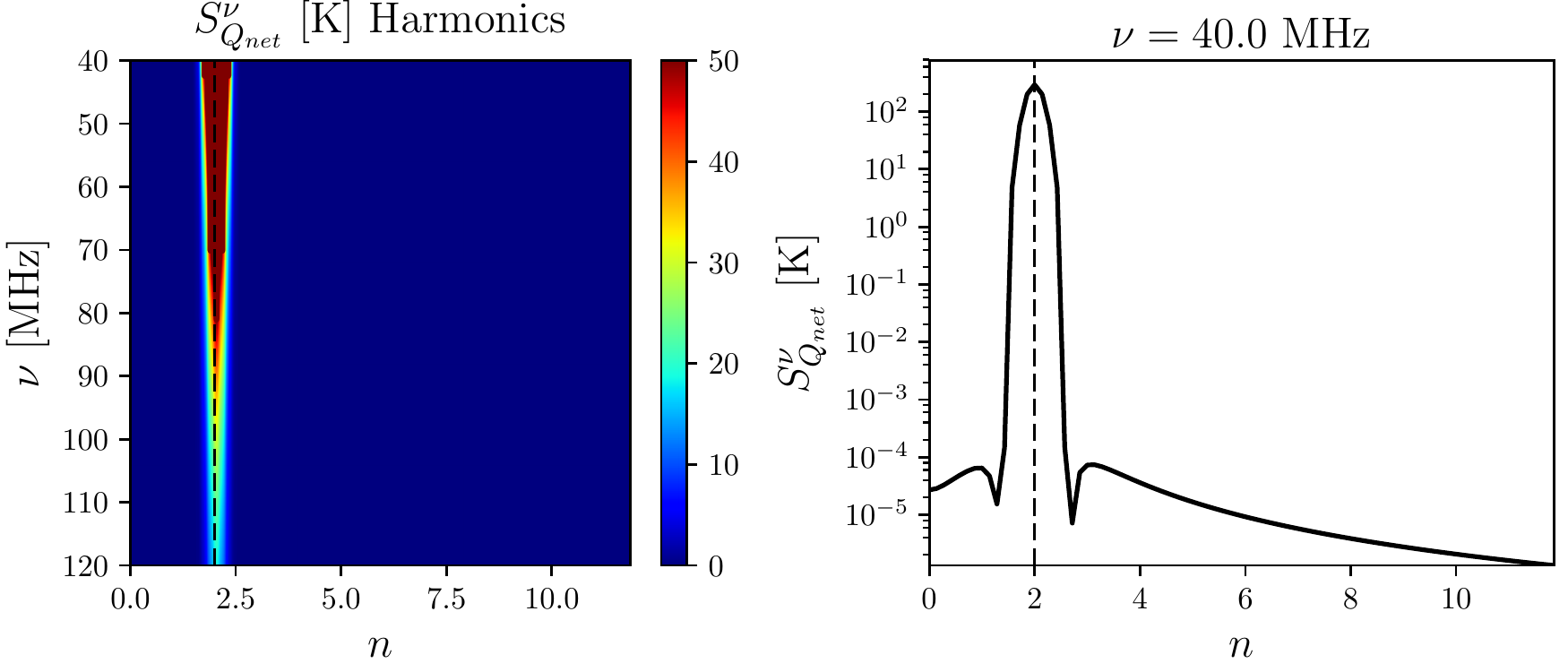} 
\caption{\emph{Left:} Similar to the total intensity measurement, FFT
  of the Stokes $Q_{net}$ helps to distinguish different harmonic
  components function of observed frequencies. \emph{Right:} The
  second-harmonic Stokes spectrum $S_{Q,2}^{\nu}$ can be constructed
  by assembling the magnitude at $n=2$ across the band of
  interest. This resulting spectrum contains only the foreground but
  not the background 21-cm
  signal.}\label{fig:stokes_Q_harmonic_gauss_circ_haslam_paper}
\end{figure}

\begin{figure}
\centering 
\includegraphics[scale=.59]{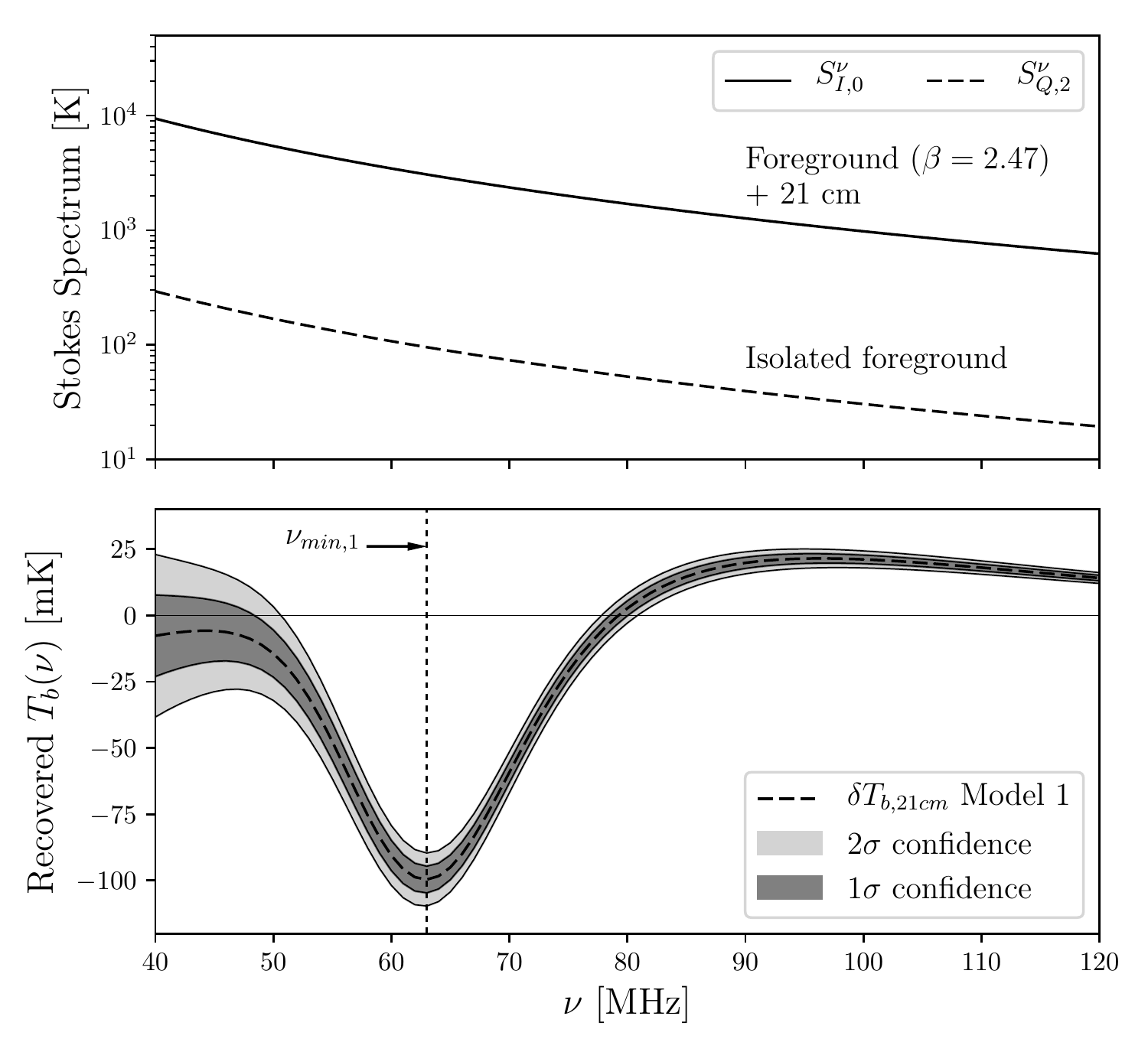}
\caption{\emph{Upper:} The Stokes spectrum $S_{Q,2}^{\nu}$ (dashed
  curve) is shown to have a spectral index identical to the input
  value $\beta$ of 2.47. By scaling and subtracting this spectrum from
  the total intensity spectrum $S_{I,0}^{\nu}$ (solid curve), the
  background 21-cm signal can be retrieved. \emph{Lower:} $1\sigma$ and
  $2\sigma$ confidence levels of the extracted 21-cm spectrum is
  compared to the input ARES spectrum for Model 1 (dashed curve). The
  primary uncertainty in this result is due to error in estimating
  $\nu_{min,1}$ (vertical dashed line) for the first derivative of
  Model 1 to compute the scaling factor $1/B_1$ of
  Equation \eqref{eq:scaling_coeff1}.}
\label{fig:stokes_extracted_spectra_gauss_circ_haslam_paper}
\end{figure}

For a circular beam, the same sky region is observed continuously over
time as the foreground revolves about the NCP, hence the
$I_{net}^{\nu}(t)$ is constant at each frequency as shown in
\autoref{fig:stokes_I_drift_scan_gauss_circ_haslam_paper}. This
translates to a zero-frequency component in its periodogram as shown
in \autoref{fig:stokes_I_harmonic_gauss_circ_haslam_paper}. In
contrary, the linear net polarization represented by the Stokes
$Q_{net}^{\nu}(t)$ is a sinusoidal function with angular frequency
that equals twice the sky's revolution rate, or two cycles per 24
sidereal hours, as illustrated in
\autoref{fig:stokes_Q_drift_scan_gauss_circ_haslam_paper}. Subsequently,
its periodogram consists of only a harmonic at $n=2$ in
\autoref{fig:stokes_Q_harmonic_gauss_circ_haslam_paper}. This second
harmonic only arises from the anisotropy in the foreground map. Hence
by constructing a second-harmonic spectrum $S_{Q,2}^{\nu}$ in a manner
similar to $S_{I,0}^{\nu}$, we obtain a replica of the foreground
spectrum, free from any isotropic background signal. This is confirmed
by calculating the spectral index $\beta$ recovered from the
$S_{Q,2}^{\nu}$, which is found to be identical to the input $\beta$
of 2.47 in
\autoref{fig:stokes_extracted_spectra_gauss_circ_haslam_paper}.

\begin{figure}
\centering 
\includegraphics[scale=.59]{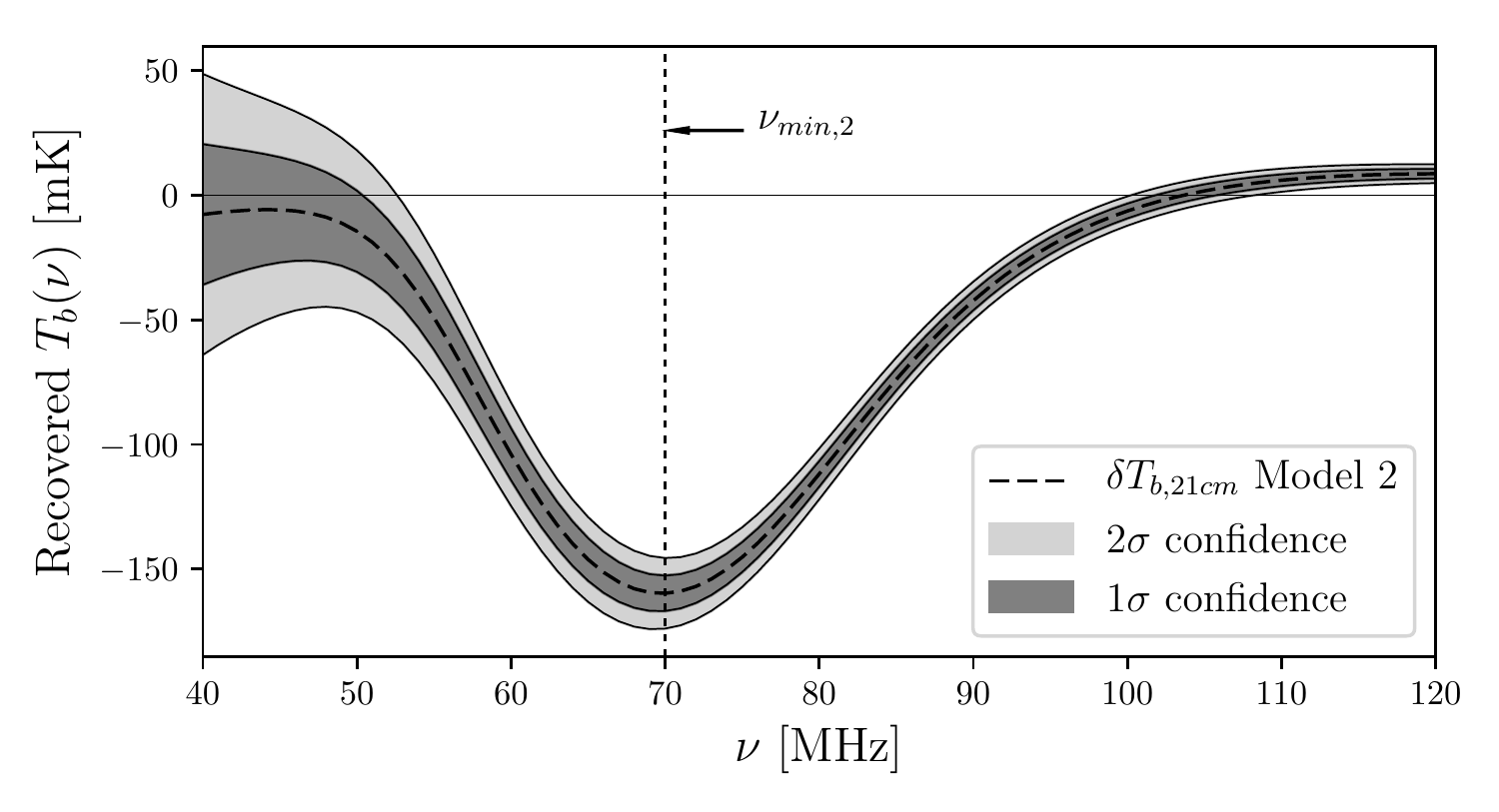}
\caption{$1\sigma$ and $2\sigma$ confidence levels of the extracted
  21-cm spectrum is compared to the input ARES spectrum for Model 2
  (dashed curve). The primary uncertainty in this result is due to
  error in estimating $\nu_{min,2}$ (vertical dashed line) for the
  first derivative of Model 2 to compute the scaling factor $1/B_1$ of
  Equation \eqref{eq:scaling_coeff1}.}
\label{fig:stokes_extracted_spectra_gauss_circ_haslam_model2_paper}
\end{figure}

In this ideal scenario, where instrumental systematics and other
contaminations can be removed, the total power spectrum is identical
to $S_{I,0}^{\nu}$ so $A_1$ is unity with zero offset $A_0$. Since
$S_{Q,2}^{\nu}$ is a scalable replica of the the foreground spectrum,
only the scaling coefficient $B_1$ is needed and no offset value $B_0$
is expected. After scaling the $S_{Q,2}^{\nu}$ up to the
$S_{I,0}^{\nu}$ with $B_1$ as in Equation \eqref{eq:scaling_coeff1}, the
foreground spectrum is subtracted to reveal the underlying global
21-cm model with Equation \eqref{eq:T21cm_extract1}, as shown in
\autoref{fig:stokes_extracted_spectra_gauss_circ_haslam_paper} and
\autoref{fig:stokes_extracted_spectra_gauss_circ_haslam_model2_paper}. 

In this simulation, we assume that the Stokes measurements have been
integrated long enough to achieve a minimal measurement precision and
the only remaining error in the extracted 21-cm model is the
uncertainty $\sigma_{AB}$ from determining the scaling factor
$A_1/B_1$, or $1/B_1$ in this case. Details on error propagation is
elaborated in following section.

\section{Implementation Aspects}
\label{sec:implementation_practicality}
\begin{figure}
  \includegraphics[scale=0.375]{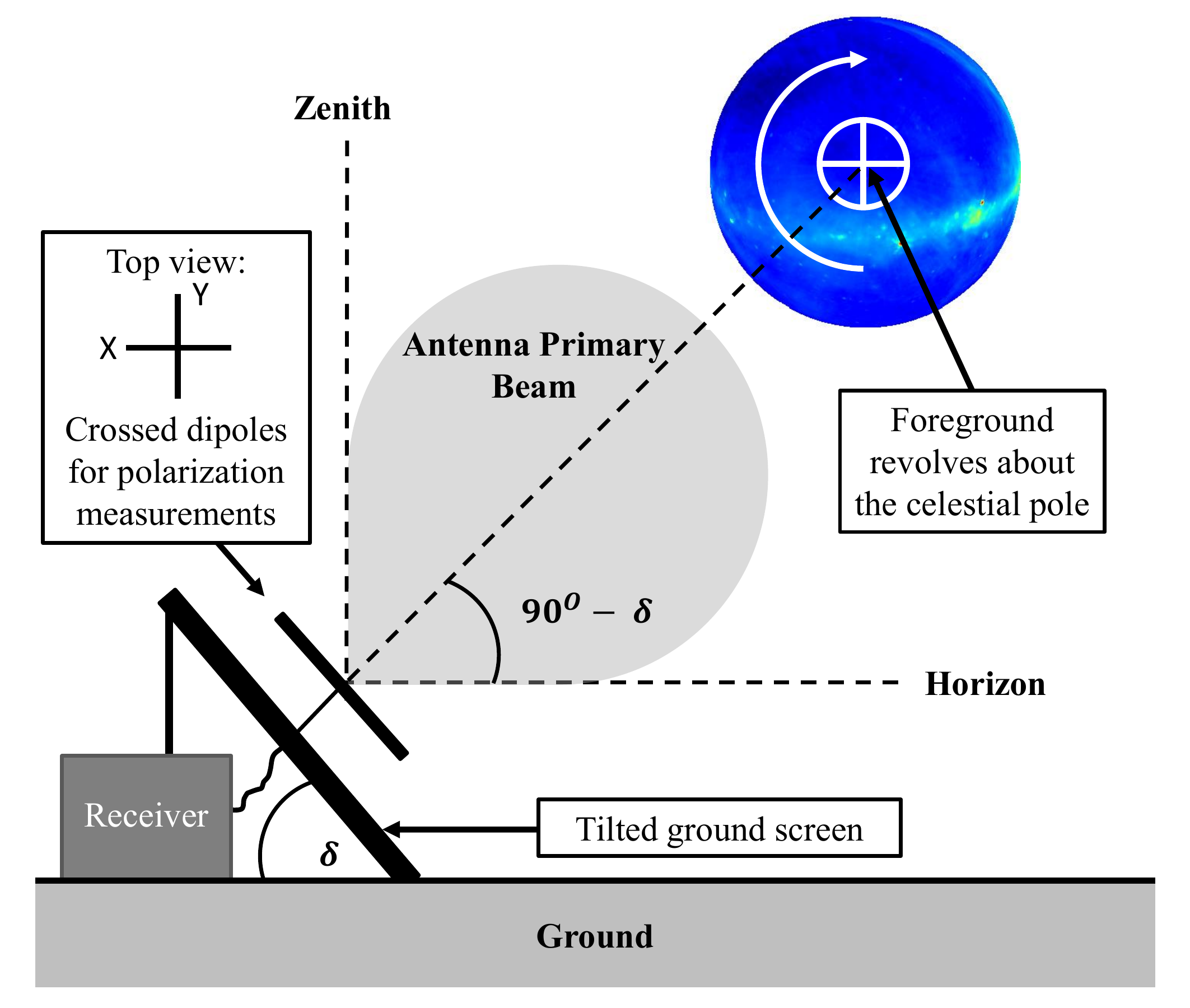} 
  \caption{Schematic for the ground-based polarimeter, which consists
    of a pair of tilted crossed dipoles pointing at a celestial
    pole. The primary differences between this polarimetric approach
    and other global 21-cm experiments are full-Stokes measurements
    and its configuration to separate the foreground from the static
    background through its own dynamic characteristic. The tilt
    angle $\delta$ corresponds to the observer's
    latitude.}\label{fig:ctp_pointing_schematic_v2}
\end{figure}

We have demonstrated that, with idealized circular Gaussian beams, the
second-harmonic spectra originated from the dynamic component of the
projection-induced polarization can be used to track the foreground
spectrum, isolated from the static 21-cm background. In this section,
we assess some of challenges when implementing such polarimetric
approach in practice. A general layout of implementing such instrument
on the ground consists of a pair of crossed dipoles being tilted such
that its boresight is pointing toward a celestial poles as illustrated
in \autoref{fig:ctp_pointing_schematic_v2}.

\subsection{Foreground subtraction error propagation}
\label{sec:foreground_subtraction_error}
As mentioned in \autoref{sec:foreground_subtraction}, the dominant
sources of error originate from estimating the derivatives of the
Stokes spectra as well as determining the $\nu_{min}$ at which the
first derivative of absorption feature in the background signal
vanishes. Because both errors are functions of the frequency
resolution, the confidence levels on the extracted Model 1 and Model 2
as shown above are attributed to the resolution chosen for this
simulation, i.e., $\Delta\nu = 1$ MHz.

Assuming the absorption feature exists at $\nu_{min}$ within the
instrument's passband, but the estimated $\nu_{min}' = \nu_{min} \pm
\Delta\nu$ with error $\Delta\nu$ is used for Equation
\eqref{eq:scaling_coeff0}, then a non-zero ${\rm d}\delta
T_{b,21cm}/{\rm d}\nu$ is unknowingly assumed to be zero when
computing $A_1/B_1$ with Equation \eqref{eq:scaling_coeff1}. In fact,
when evaluating Equation \eqref{eq:scaling_coeff0} at $\nu_{min}' \neq
\nu_{min}$, we obtain a non-zero value on the right-hand side (RHS) in
the following equation,
\begin{equation}
  \frac{A_1}{B_1} - \frac{{\rm d}S_{I,0}^{\nu}}{{\rm d}\nu} \left(\frac{{\rm
      d}S_{Q,2}^{\nu}}{{\rm d}\nu}\right)^{-1} = 
  A_1k_B\frac{{\rm d}\delta T_{b,21cm}(\nu)}{{\rm d}\nu} \left(\frac{{\rm
      d}S_{Q,2}^{\nu}}{{\rm d}\nu}\right)^{-1}
  \label{eq:scaling_coeff2}
\end{equation} 
Since the frequency resolution of the simulations is set to 1 MHz, the
first derivative of $\delta T_{b,21cm}$ does not equal to zero exactly
at the estimated $\nu_{min}'$. So we estimate the $\sigma_{AB}$ to be
the unknown difference between the values of $A_1/B_1$ at the correct
$\nu_{min,1}$ and at the $\nu_{min}'$. This error is essentially the
term on the RHS but we are not using the derivative of our input Model
1 directly.

By applying a basic error propagation to Equation
\eqref{eq:T21cm_extract1}, the uncertainty of the extracted 21-cm
signal $\sigma_{21cm}$ can be estimated as,
\begin{equation}
  \begin{aligned}
    \sigma_{21cm}(\nu) = & \frac{1}{k_BA_1} \biggl[\sigma_{I,0}^2(\nu) + 
      \left(\frac{A_1}{B_1}\right)^2\sigma_{Q,2}^2(\nu)\ + \\
      & (S_{Q,2}^{\nu}-B_0)^2\sigma_{AB}^2 \biggr]^{1/2}
  \end{aligned}
  \label{eq:21cm_error}
\end{equation}
where $\sigma_{I,0}$ and $\sigma_{Q,2}$ are the measurement
uncertainties of the Stokes spectra $S_{I,0}^{\nu}$ and
$S_{Q,2}^{\nu}$ defined by the radiometer equation in Equation
\eqref{eq:radiometer_eq}. In the simulations above, we assume the
measurement uncertainty of the Stokes spectra have been reduced to a
minimal level after longterm integration, so the overall uncertainty
of the extracted signal which defines the confidence levels reduces
to,
\begin{equation}
  \sigma_{21cm}(\nu) = (S_{Q,2}^{\nu}-B_0)\sigma_{AB} = S_{Q,2}^{\nu}\sigma_{AB}
  \label{eq:21cm_error_AB}
\end{equation}

\subsection{Searching for $\nu_{min}$}
\label{sec:foreground_subtraction_error}
\begin{figure}
\centering 
\includegraphics[scale=.59]{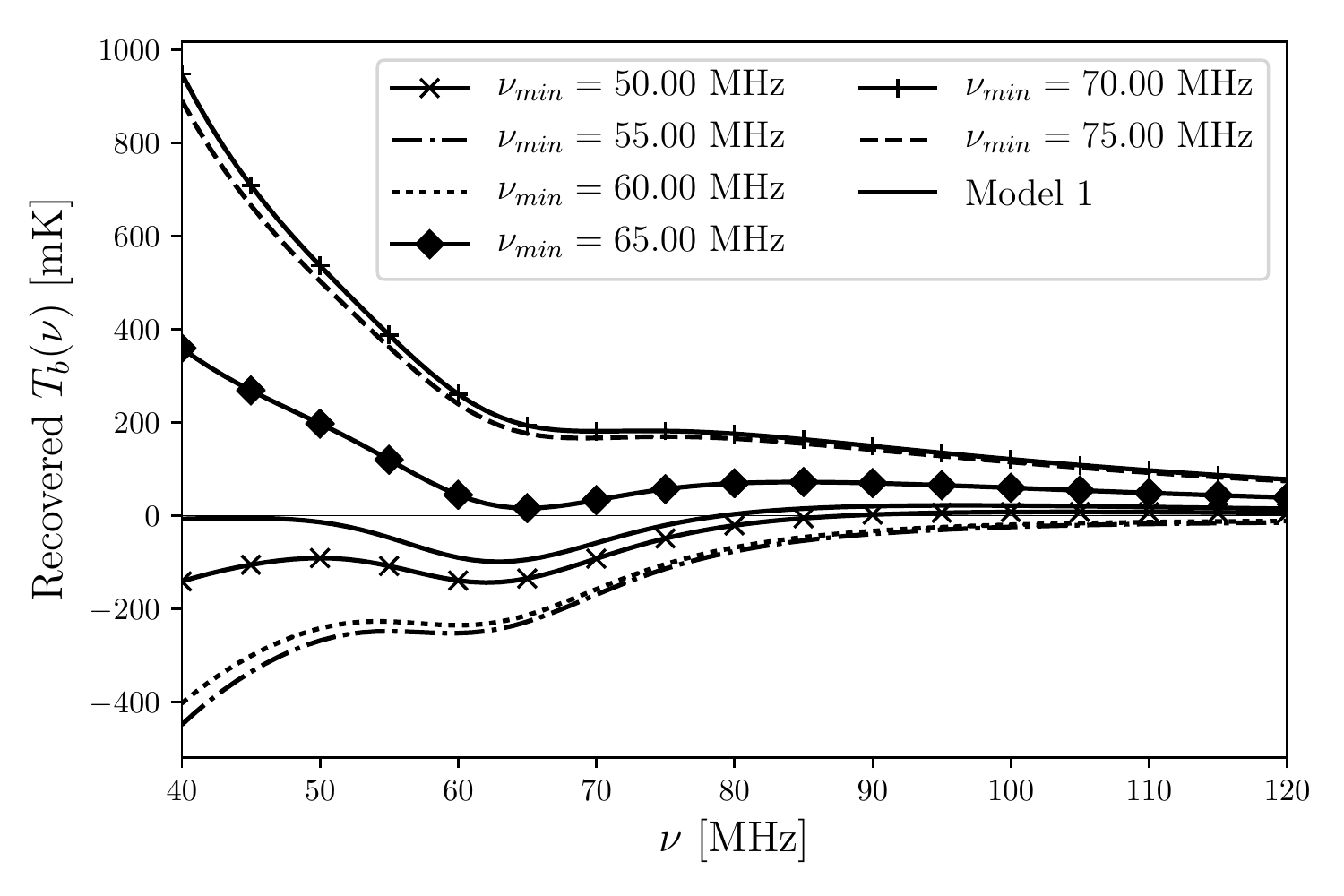}
\caption{Illustration on how the scaling factor $A_1/B_1$ can be
  refined iteratively by a blind search for frequency $\nu_{min}$ of
  the global 21-cm absorption feature, at which its first derivative
  equals to zero. Shown here is a series of recovered candidates for
  the input 21-cm model spectrum. Each of the curve is estimated by
  scaling and subtracting the second-harmonic spectrum from the
  total-power spectrum using a $A_1/B_1$ factor computed at an assumed
  $\nu_{min}$ as shown in the legend, without using any information
  from the input Model 1.
}\label{fig:stokes_spectra_scale_compare_gauss_circ_haslam_paper}
\end{figure}

Except for the derivatives of the Stokes spectra on the left-hand side
(LHS) of Equation \eqref{eq:scaling_coeff2}, the scaling factor and
the background signal's first derivative are unknowns. As an observer,
we are simply searching for a frequency of the absorption feature in
the background so that the second terms on the RHS vanishes even
without knowing where it is. We here propose a blind search to
determine an optimal $\nu_{min}$ during the foreground
subtraction. After an initial foreground subtraction by choosing a
random frequency as $\nu_{min}$ in the passband, the process of
scaling and subtracting $S_{Q,2}^{\nu}$ from $S_{I,0}^{\nu}$ can be
refined iteratively (ideally from one end of the passband to the
other) until potential spectral structures may appear in the residual
spectrum, as illustrated in
\autoref{fig:stokes_spectra_scale_compare_gauss_circ_haslam_paper}.

\begin{figure}
\centering 
\includegraphics[scale=.59]{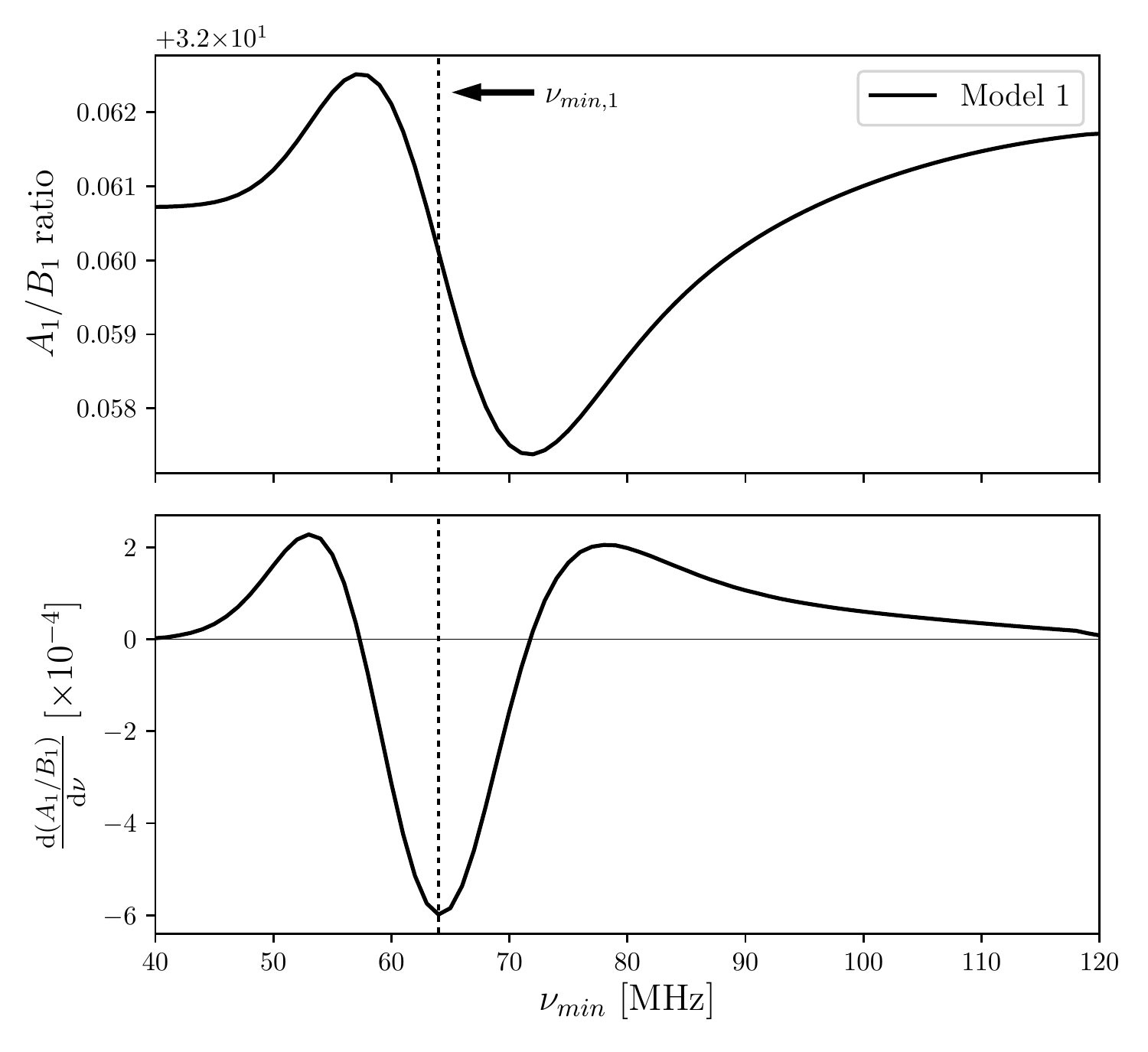}
\caption{\emph{Upper:} This illustrates how the factor
    $A_1/B_1$ varies a function of the chosen $\nu_{min}$ during the
    blind search procedure. Since each chosen $\nu_{min}$ is used by
    assuming it is the frequency of the absorption feature, $A_1/B_1$
    represents the unknown error contributed by the term on the RHS of
    Equation \eqref{eq:scaling_coeff2}. Hence it is proportional to
    the negative value of the global 21-cm spectrum's first
    derivative.  \emph{Lower:} Since the magnitude of a second
    derivative helps to distinguish a local minimum from a global one,
    taking the first derivative of $A_1/B_1$ respect to $\nu_{min}$
    provides an additional information to allow the iterative process
    to converge onto the desired $\nu_{min,1}$ (as shown in the dashed
    line).}\label{fig:stokes_spectra_scale_vs_nu_min_gauss_circ_haslam_paper}
\end{figure}

Without the background signal and its absorption feature as a priori
information, the iterative scaling process only converges when it
locates a zero crossing in the global 21-cm spectrum's derivative. By
assuming that each of the chosen $\nu_{min}$ can potentially be the
desired value, we can simply calculate $A_1/B_1$ by ignoring the extra
term on the RHS of Equation
\eqref{eq:scaling_coeff2}. \autoref{fig:stokes_spectra_scale_vs_nu_min_gauss_circ_haslam_paper}
shows how $A_1/B_1$ varies as a function of $\nu_{min}$. By the
formulation of Equation \eqref{eq:scaling_coeff2}, it is not
surprising that the curve is proportional to the negative value of
Model 1's first derivative in
\autoref{fig:model_signal_derivatives_paper}. More importantly, since
local and global minima can be distinguished by the magnitude of the
second derivative, we can determine the desired $\nu_{min,1}$ simply
by taking the first derivative of $A_1/B_1$ respect to $\nu_{min}$, as
shown in
\autoref{fig:stokes_spectra_scale_vs_nu_min_gauss_circ_haslam_paper},
without any knowledge of the background signal.

It is worth noting that the approach presented here only applies to an
idealized instrument with optimized calibration, which is free from
any spurious contamination and unwanted distortion. Presence of
realistic instrumental systematics and other measurement uncertainties
can complicate this foreground subtraction procedure. More
sophisticated estimation algorithms, which utilize theoretical
constraints from the 21-cm physics and instrument models in terms of
the Bayesian statistics such as the Monte Carlo Markov Chain
\citep[MCMC, ][]{harker2012mcmc, mirocha2015interpreting} and SVD, are
needed in parallel. This will be investigated in future work.

\subsection{Stokes spectra measurement sensitivity}
\label{sec:spectra_sensitivity}
In the full sky simulations above, the Stokes spectra are assumed to
have been integrated long enough to achieve an optimal measurement
precision such that only $\sigma_{AB}$ dominates the overall error in
extracting the 21-cm models as shown in
\autoref{fig:stokes_extracted_spectra_gauss_circ_haslam_paper} and
\autoref{fig:stokes_extracted_spectra_gauss_circ_haslam_model2_paper}. By
considering only the contribution of thermal noise after other
systematics have been corrected and treating $\sigma_{AB}$ as a
second-order effect, we estimate the required integration time to
achieve a $\sigma_{I,0}/k_B$ of at least 10 part per million (ppm) of
the expected sky noise temperature with the radiometer equation.

The generic radiometer equation in Equation \eqref{eq:radiometer_eq}
suggests the measuring uncertainty (or minimal detectable temperature
change) at each frequency channel $\Delta T_{min}(\nu)$ is dictated by
the systematic noise temperature $T_{sys}(\nu)$ and decreases as a
function of the overall integration time $\tau_{int}$ and the
available bandwidth $\Delta\nu$ in the absence of any receiver gain
variations.
\begin{equation}
  \Delta T_{min}(\nu) = \frac{T_{sys}(\nu)}{\sqrt{\tau_{int}\Delta\nu}}
  \label{eq:radiometer_eq}
\end{equation}
From the simulation, at 40 MHz, the total sky temperature is about
9400 K. By assuming a receiver noise temperature $T_{rx} = 300$ K, the
total system noise temperature $T_{sys} \sim 9700$ K. For different
bandwidth value, using the radiometer equation, a rudimentary
estimates of integration time as calculated.
\begin{table}[!htbp]
  \caption{Integration time estimates for different spectral resolution} \label{tab:integration_time} \centering
  \begin{tabular}{ l | l } 
    \hline  
    \hline
    $\Delta\nu$ \T\B\ $[{\rm Hz}]$ & $\tau_{int}\ [{\rm Hr}]$\\
    \hline
    $1.0 \times 10^{4}$\T & $2.46 \times 10^4$\\
    $5.0 \times 10^4$ & $4.92 \times 10^3$\\
    $1.0 \times 10^5$ & $2.46 \times 10^3$\\
    $5.0 \times 10^5$ & $4.92 \times 10^2$\\
    $1.0 \times 10^6$ & $2.46 \times 10^2$\\
    \hline                                   
  \end{tabular}
\end{table}
As shown in \autoref{tab:integration_time}, a reasonably short
integration time is needed for $\Delta\nu > 100$ kHz to achieve the
required sensitivity under the given assumptions.

\subsection{Chromatic beam effects}
\label{sec:effect_beam}
In global 21-cm experiments, it is desirable to have a broadband
antenna with relatively low variations in the beam patterns as a
function of frequency. Since the intrinsic physical size and
surrounding environment of a resonant antenna structure determine the
available bandwidth and its efficiency, it is unavoidable to have an
antenna radiation pattern $F(\theta,\phi,\nu)$ varies as a function of
frequencies. The beam patterns are related to the antenna directive gain
which determines the antenna's effective FOV. Hence, the effective
collecting area is altered by the frequency dependent antenna gain,
which perturbs the total observable power. This chromatic dependence
of the antenna beams corrupts the spectrally-smooth foreground by
introducing unwanted spectral structures into the sky measurements
\citep{vedantham2013chromatic, bernardi2015foreground,
  mozdzen2016limits}.

The projection-induced polarization relies on the overall foreground
anisotropy to a produce a net Stokes vector, but the level of anisotropy
depends on the beams' effective FOV. Hence the variations in the
antenna gain with frequency will perturb the measured net Stokes
vector. The spectral shape of the constructed Stokes spectra, like
$S_{I,0}^{\nu}$ and $S_{Q,2}^{\nu}$, are also therefore
distorted. Depending on the magnitude of these gain variations, the
resulting unwanted spectral structures can be as large as, if not
greater than, the predicted global 21-cm signal.

In addition to the perturbed FOV due to the antenna gain variations,
geometry of the beam patterns can also affect the spectral shape of
the Stokes spectra. In practice, it is difficult to design an antenna
with a circular beam across the entire passband. Due to the intrinsic
characteristics of a dipole resonator, a certain level of beam
elongation (or ellipticity) is often present in the beam patterns.

\begin{figure}
\centering 
\includegraphics[scale=.5]{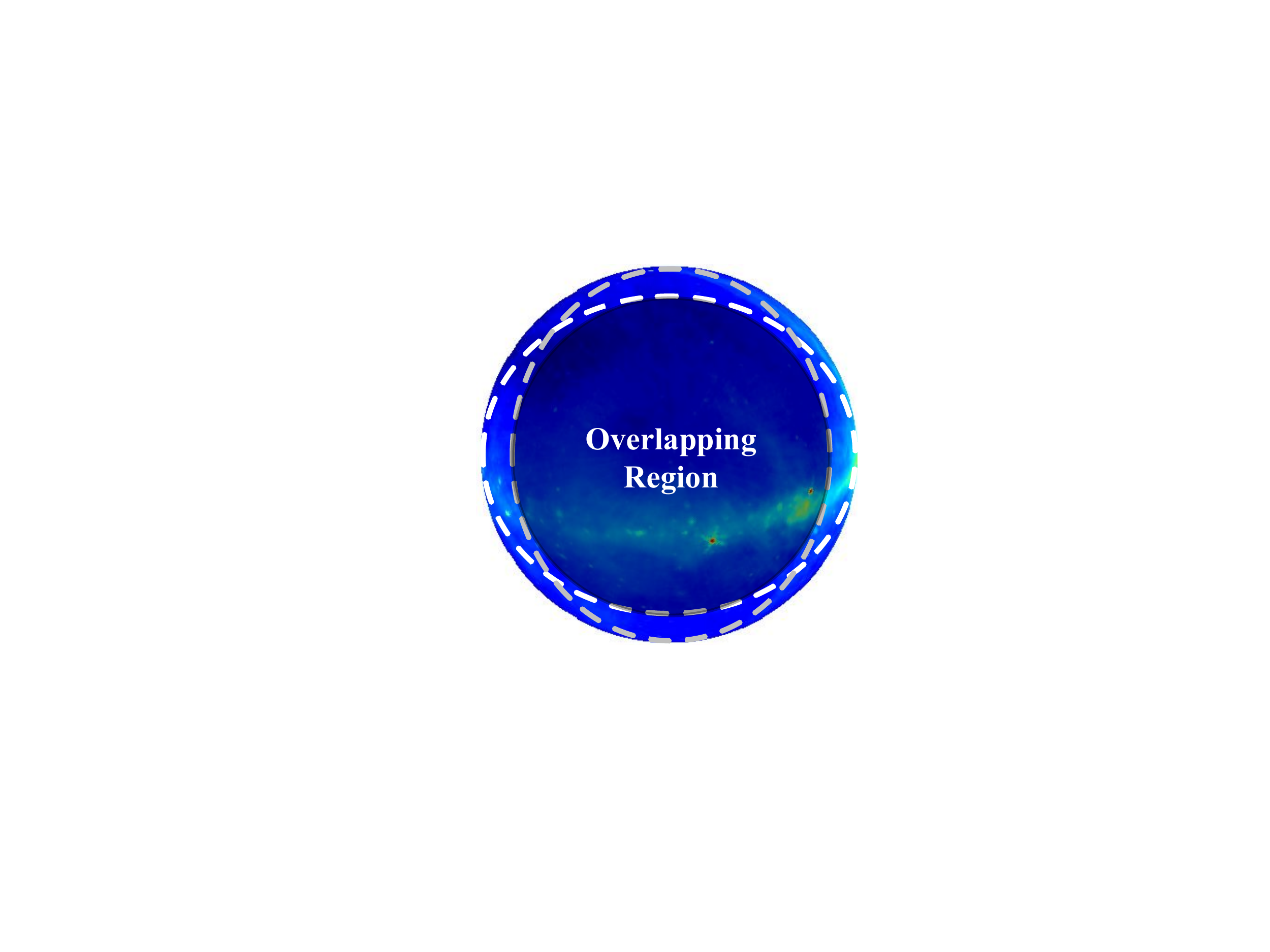}
\caption{Illustration of the overlapping sky region covered by two
  elongated beams from the crossed dipoles. The overlapping sky region
  can be observed continuously by both antenna, but not the exterior
  regions.}\label{fig:elongated_beam_illustration}
\end{figure}

If the crossed dipoles have circular beams, they cover the same sky
region the entire time so that the cross-correlation between the $X$
and $Y$ oriented antenna are identical. However, as illustrated in
\autoref{fig:elongated_beam_illustration}, when the dipoles have two
orthogonal elongated beams, only the overlapping region of the two
beams covers the same sky at the entire time. This overlapping region
is maximized when the two beams aligned in the case of circular
ones. The sky portion outside of this overlapping region is
  only sampled whenever it is aligned with the dipoles. This
  contributes fourth-order harmonic component ($n=4$) into the Stokes
  parameters $I$ and $I_p$. As the beam elongates differently from
frequency to frequency, this introduces spectral variations to the
Stokes measurements, thus to the foreground spectrum.

These chromatic distortions can compromise the detection sensitivity
unless the antenna beams are perfectly known. One of the probable ways
to mitigate this effect is to reference the measurements to numerical
antenna beam models. Careful incorporation of the antenna structure
design and environment allows sophisticated electromagnetic simulation
software, such as the Computer Simulation Technology (CST) or the High
Frequency Structural Simulator (HFSS), to produce high-fidelity
antenna beam models. By bootstrapping these beam models to the
measurements, the spectral-dependent beam distortions can be
mitigated, if not removed. Rudimentary analysis suggests that as long
as the antenna beam pattern can be constrained with accuracy down to
few tens ppm, the chromatic effects on the sky measurements can
potentially be corrected to a desired level. In principle, the
fourth-order harmonic is tracking the beam elongation and potentially
can be used as feedback information to the beam modeling and
calibration. However, the full potential of such beam correction using
numerical beam model will be explored in future work.

\subsection{Ground effects on the antenna beams}
\label{sec:effect_ground}
Our simulations assume a pair of crossed-dipole antennas located at
the GNP (at latitude $\phi_{\oplus} = 90^{\circ}$) to have a full FOV
the northern sky centered at the NCP. For practical and logistical
reasons, such an instrument can only be deployed at locations of
latitude between $0^{\circ} < \phi_{\oplus} < 90^{\circ}$, where the
Equator is at $\phi_{\oplus} = 0^{\circ}$. The antenna will be
pointing toward the NCP at a tilt angle $\delta$ which is a function
of the observer's latitude.

\begin{figure}
\centering 
\includegraphics[scale=.375]{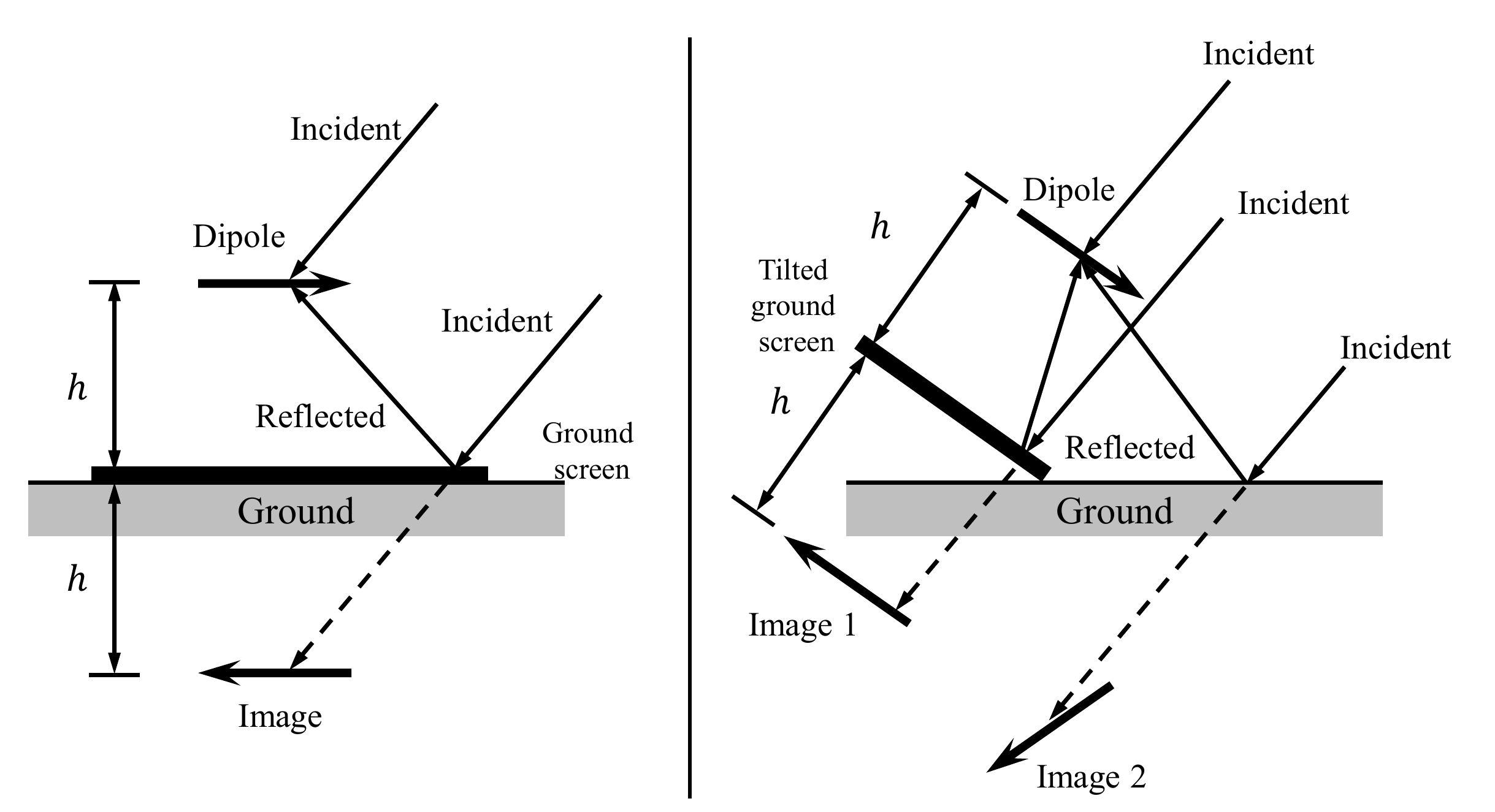}
\caption{\emph{Left:} There is a single image when the horizontal
  dipole is above the ground plane at a height of $h$. \emph{Right:}
  Multiple images are present when the dipole are tilted relative to
  the flat ground. This introduces unwanted interferometric fringes
  that distort the smooth dipole
  beams.}\label{fig:ground_image_comparison}
\end{figure}

\begin{figure}
\centering 
\includegraphics[scale=.475]{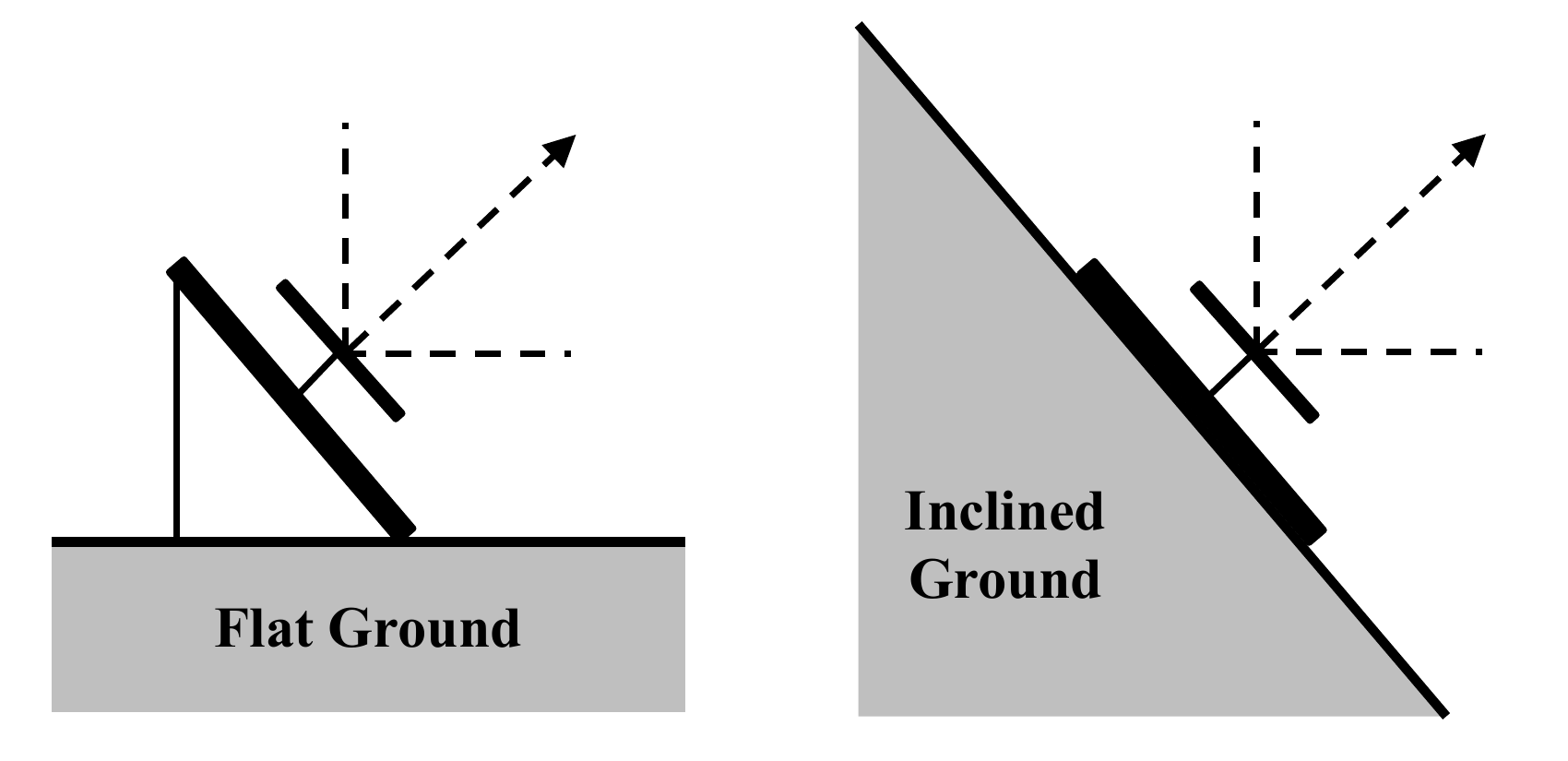}
\caption{To mitigate the ground effects on the beams, the tilted
  dipole can be situated on an inclined ground (\emph{Right}) with a
  slope similar to the tilt angle $\delta$ instead of a flat ground
  (\emph{Left}), depends on the observer's
  latitude.} \label{fig:ctp_ground_placement_comparison}
\end{figure}

From simple image theory formulation, it is shown that the farfield
antenna beams for a horizontal dipole above a finite ground plane at
height $h$ are smooth pattern resembling the Gaussian beams
\citep{balanis2016antenna}. However, as the antenna and its ground
plane are being tilted to point at the celestial pole, instead of a
single image, multiple images of the dipole are reflected across the
ground screen as well as the actual ground as illustrated in
\autoref{fig:ground_image_comparison}. The different images from
different propagation distances result in phase delays such that
interferometric fringes are introduced into the beams. These ground
effects can be mitigated by situating the antenna system on a slope
instead of on the flat ground as in
\autoref{fig:ctp_ground_placement_comparison}. The slope angle needs
to be consistent with the tilt angle $\delta$. Given the proposed
configuration, any potential ground effects on the antenna beam will
be present in the measured Stokes parameters as constant with time. It
is worth noting that such beam distortion due to the ground can very
likely impose a $n=1$ harmonic ,if not at higher orders ($n>4$), to
the Stokes measurement instead of at $n=2$ due to the relative
geometry of the antenna and the sky.

\subsection{Horizon obstruction}
\label{sec:effect_horizon}
\begin{figure}
\centering 
\includegraphics[scale=.45]{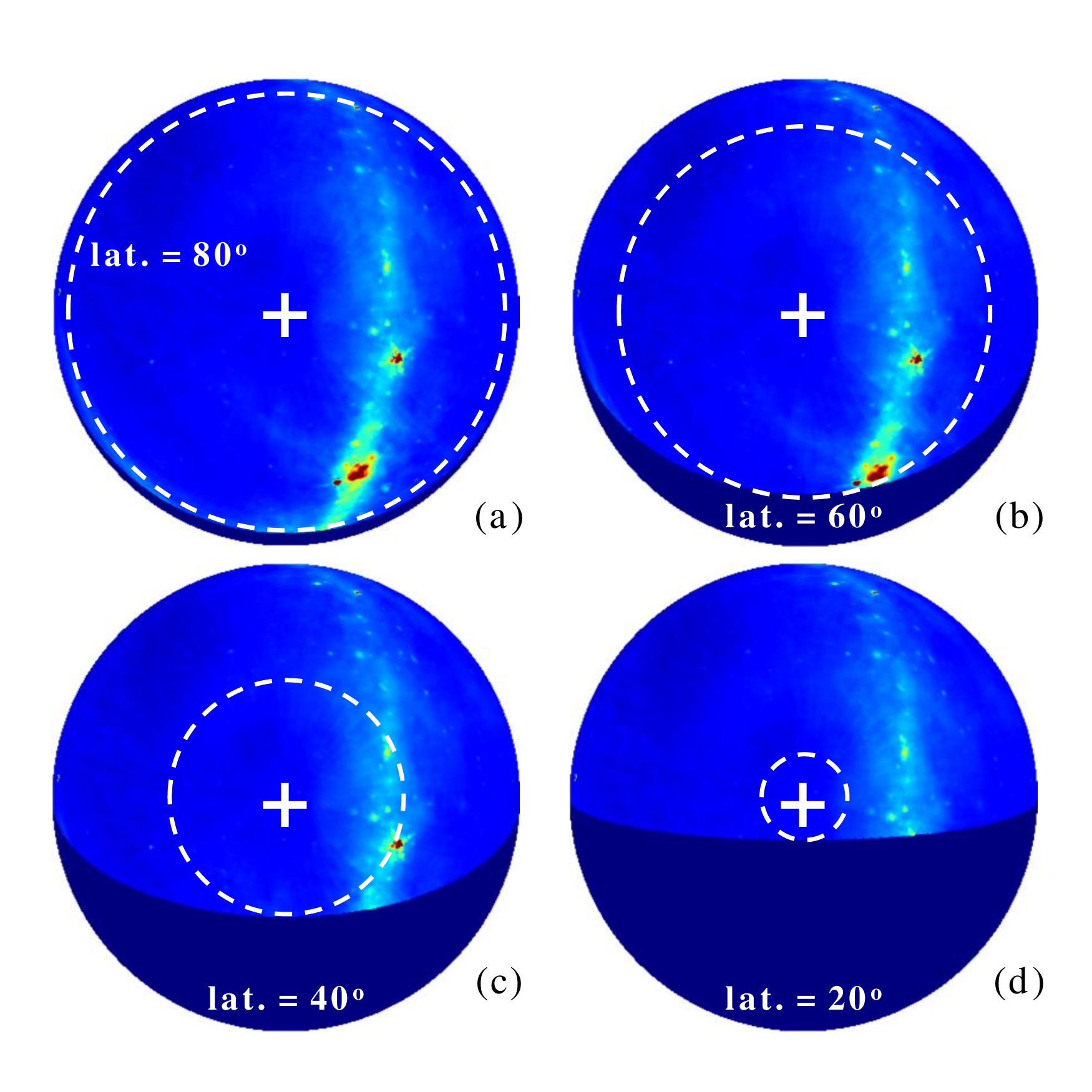}
\caption{The observable sky centered at the NCP is partially
  obstructed by the horizon as the latitude of the observation site
  decreases: (a) $\phi_{\oplus} = 80^{\circ}$, (b) $\phi_{\oplus} =
  60^{\circ}$, (c) $\phi_{\oplus} = 40^{\circ}$, and (d)
  $\phi_{\oplus} = 20^{\circ}$. As a result, the inner sky region is
  observable the entire time (inside the white dashed circle), whereas
  the outer region rises and sets once per revolution. As the inner
  sky region shrinks, so do the amplitude hence the SNR of the
  second-harmonic spectrum. A first-order harmonic is introduced to
  the linear Stokes parameters due to the rising an setting of the
  outer sky region. }\label{fig:foreground_horizon_illustration}
\end{figure}

As the latitude decreases toward the Equator, the FOV is partially
obstructed by the horizon thus the visible sky can be separated into
two parts. Only the inner region of the sky about the boresight, as
shown within the dashed circle in
\autoref{fig:foreground_horizon_illustration}, is observable the
entire time, whereas the sky region outside the dashed circle rises
and sets once per revolution.

As pointed out in previous sections, the magnitude of the second
harmonic in projection-induced polarization depends on the overall
anisotropy of the sky region which can be observed the entire time
continuously. Hence, with the presence of the horizon, the sky region
that contributes to the second harmonic in $S_{Q,2}^{\nu}$ and
$S_{U,2}^{\nu}$ is this inner sky region within the dashed circle. As
a result, smaller radii of the inner sky region at lower latitudes
imply lower accuracy of the second harmonics relative to overall
system noise. Additionally, as the outer sky region rises and sets, it
contributes additional terms to the projection-induced polarization
with angular frequency of at least once per revolution to produce a
$n=1$ harmonic. Meanwhile, the sharp cutoff of the horizon will also
introduce high-frequency components to the periodic waveforms in the
Stokes parameters which can be identified at $n>4$ in the PSD. Similar
to other effects mentioned above, although these additional harmonics
can take over the power from the second harmonic and lower its
sensitivity, their effects are attenuated because the antenna gain
decreases at larger angles away from the boresight for the outer sky
region.

By incorporating the observer's latitude, effects of the horizon
obstruction on the Stokes measurement can be characterized and
corrected if needed. A balance is also needed between site selection,
RFI prevention, SNR of the projection-induced Stokes parameters, as
well as minimal ionospheric disturbance. For example, selecting a
radio-quiet zone with a latitude range that can accommodate a minimal
horizon obstruction yet not close to the Earth's magnetic poles, at
which strong interaction between the ionosphere and cosmic rays is
well known \citep{newell2001role}. An alternative to avoid the Earth's
horizon effect is to adapt this technique into a space-based
instrument, such as what is being proposed for DARE.

\subsection{Beam pointing error}
\label{sec:pointing_error}
\begin{figure}
\centering 
\includegraphics[scale=.45]{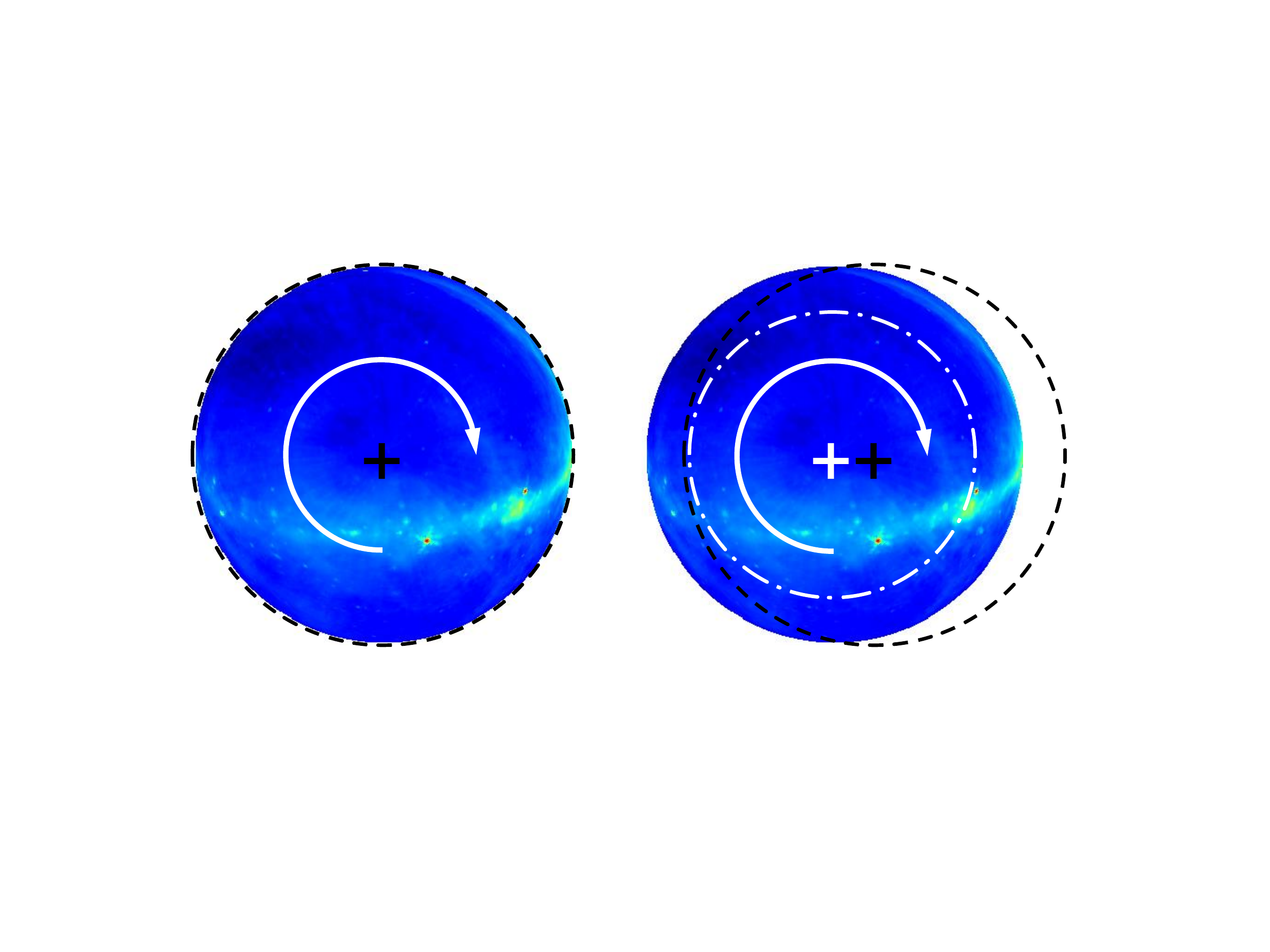}
\caption{\emph{Left:} The antenna pointing is aligned to the NCP hence
  the sky revolution is concentric to the antenna's field of
  view. \emph{Right:} Misalignment between the antenna's boresight
  (black cross) and the NCP (white cross) results in only a portion of
  the visible sky (within the white dot-dashed circle) that is
  observable continuously the entire time. The white arrow indicates
  the orientation of the sky rotation relative to the
  observer.} \label{fig:pointing_error_illustration}
\end{figure}

By design, the accuracy of measuring the second harmonic component in
the projection-induced polarization relies on the antenna pointing.
If the antenna boresight is not aligned with the celestial poles, the
foreground region centered at the celestial pole is not concentric
with the antenna beams, as illustrated in
\autoref{fig:pointing_error_illustration}. As the sky revolves around,
the foreground appears to ``wobble'' about the off-centered pointing.

Similar to the horizon obstruction, the effective sky region that
produces the $n=2$ component (within the dot-dashed circle) is reduced
which decreases the power of the second harmonic spectrum thus limits
the accuracy of the Stokes measurements. As the off-centered sky
revolves around, amplitudes and angular frequencies of the Stokes
parameters are modulated such that additional harmonics are introduced
in their PDS, for example an additional $n=4$ component can be
introduced to the measured intensity $I$ and $I_p$. These modulations
depend on the frequency-dependent beam sizes, hence they will also
cause unwanted spectral structures to appear in the measured sky
spectrum because the modulations correlate to the frequency-dependent
beam sizes. Being said, these effects can easily be mitigated by
improving the pointing accuracy. Based on our simulations, we find
that the error is acceptable for a pointing error of $1^{\circ}$ or
less relative to the celestial pole based.

\subsection{Effects of intrinsic foreground polarization}
\label{sec:effect_intrinsic_pol}
\begin{figure}
\centering 
\includegraphics[scale=.65]{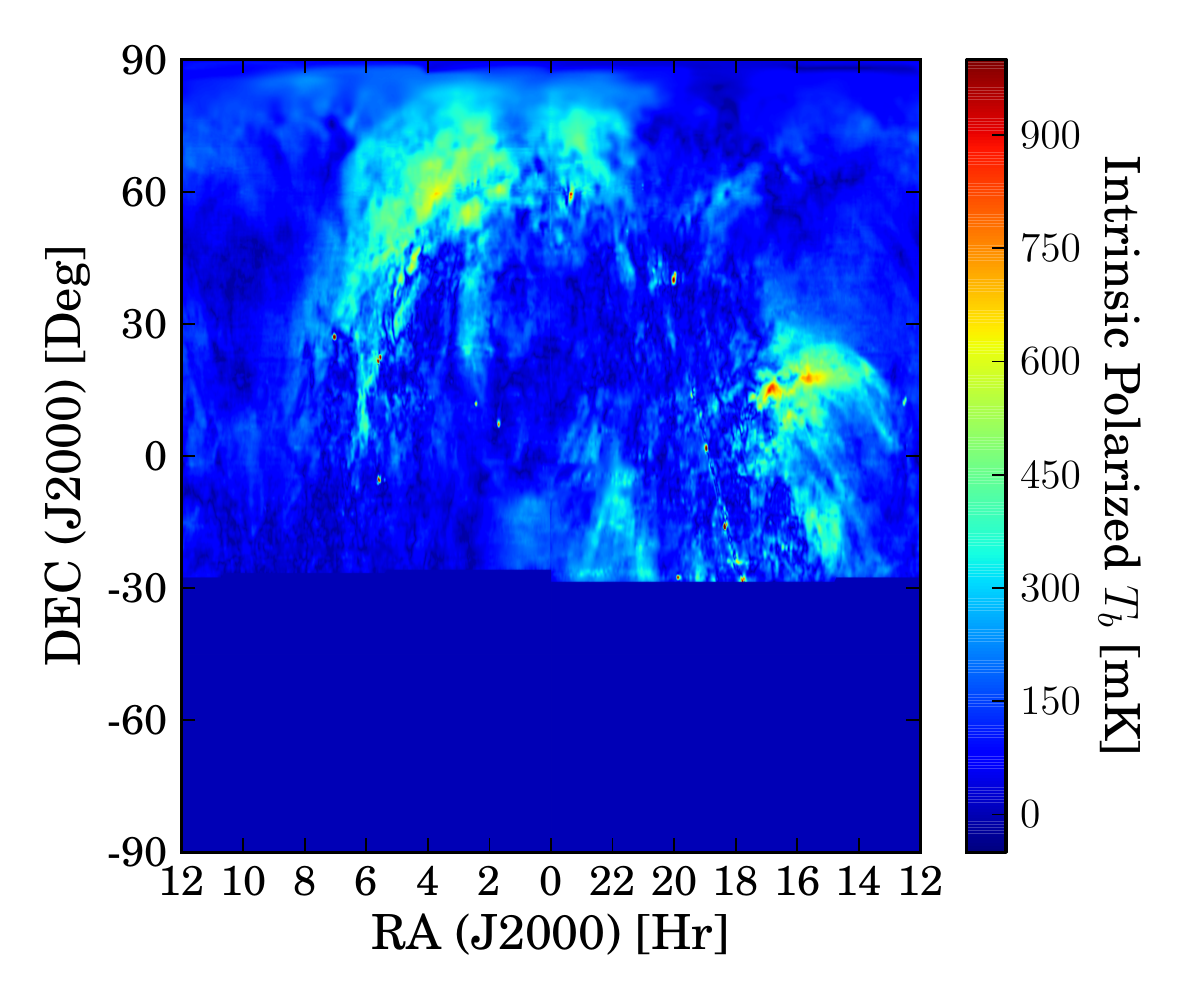} 
\caption{Intrinsic foreground polarized temperature map obtained from
  the DRAO sky survey at 1.4 GHz.}\label{fig:drao_I_QU}
\end{figure}

Synchrotron emission Galactic foreground is well known to have a
linearly polarized component. As an example, \autoref{fig:drao_I_QU}
illustrates the spatial distribution of the intrinsic foreground
polarized brightness temperature observed from the DRAO 1.4 GHz sky
survey \citep{wolleben2006absolutely}. 

As the polarized incoming signal propagates through the magnetized
interstellar medium (ISM), the polarization angle of the signal
rotates, which is known as Faraday rotation. In Gaussian cgs units,
the change in propagation phase angle is given as
\citep{wilson2009tools},
\begin{equation}
  \Delta\psi = \lambda^2{\rm RM} 
  = \frac{e^2}{2\pi m_e^2 c^2} \frac{1}{\nu^2} \int_{0}^{L}B_{\|}(l)n_e(l){\rm d}l
  \label{eq:faraday_rotation}
\end{equation}
where the rotation measure ${\rm RM}$ is defined to be the integration
of the electron number density $n_e(l)$ with the ISM magnetic field
component that is parallel to the direction of propagation $B_{\|}(l)$
at each distance increment ${\rm d}l$ along the path of the medium
with thickness of $L$. The other physical constants are the charge of
an electron $e$, electron mass $m_e$, and the speed of light in vacuum
$c$.

Due to irregular distribution of electron density in the ISM as well
as the Galactic magnetic field, the frequency-dependent Faraday
rotation of the incoming polarized foreground emission will be
different across the band. Hence, unwanted spectral structures can
also be introduced to the measured spectrum. In fact, the
intrinsic foreground polarization has been identified as the main
cause of the polarization leakage which complicates the instrument
calibration in the interferometric EoR experiments
\citep{jelic2014initial, jelic2015linear, asad2015polarization}.

\begin{figure}
\centering
\includegraphics[scale=.5]{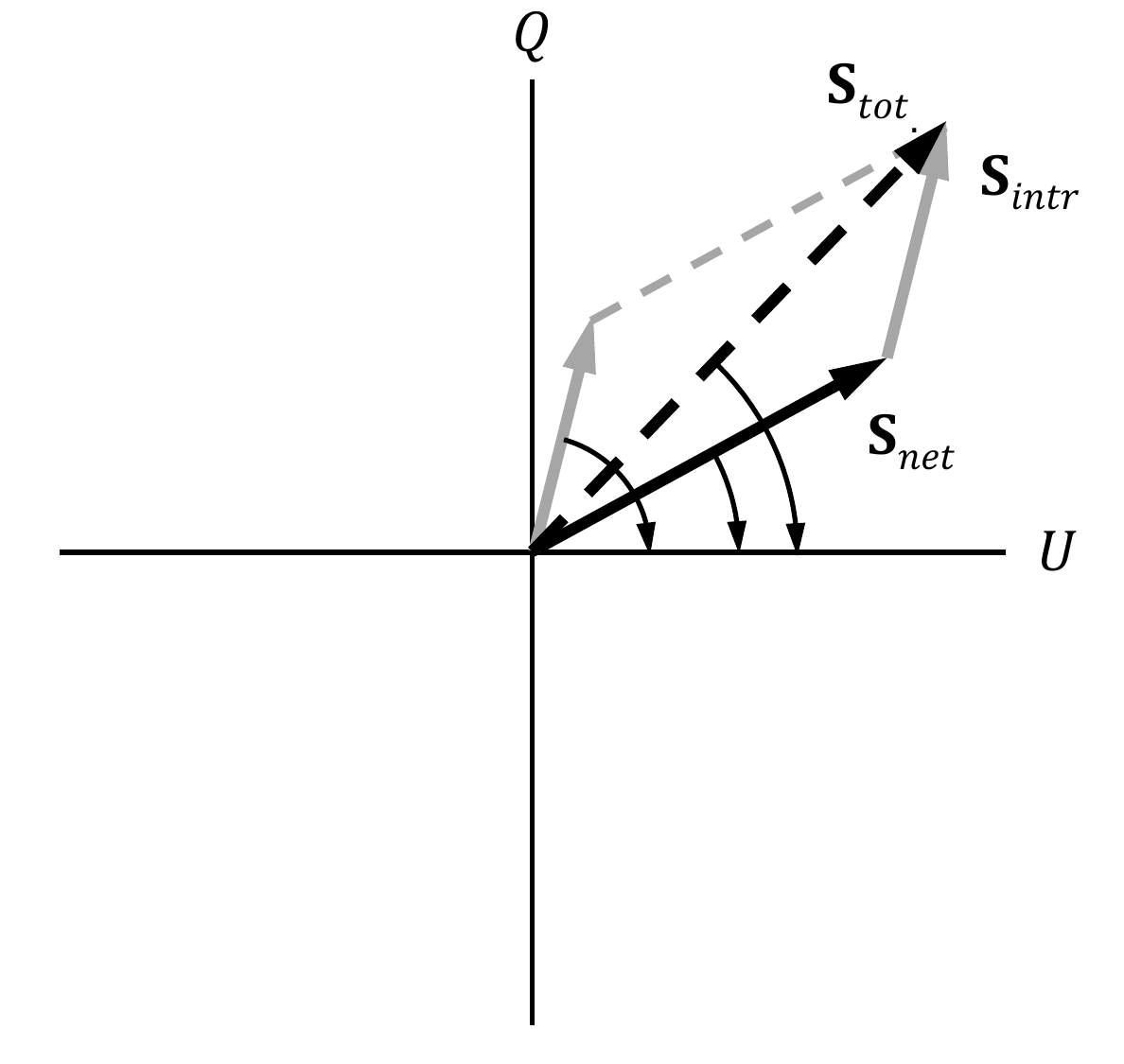}
\caption{Intrinsic foreground polarization $\bold{S}_{intr}$ (solid
  gray) contributes to the projection-induced polarization
  $\bold{S}_{net}$ (solid black) as a vector summation. Since the
  intrinsic polarization follows the foreground rotation, the
  resulting total measured polarization $\bold{S}_{tot}$ (dashed
  black) also contains a second harmonic component which can be used
  to constrain the foreground
  spectrum.}\label{fig:intrinsic_fg_polarization_vector_illustration}
\end{figure}

A single polarization dipole system in conventional global experiment
may be insensitive to linearly polarized portion of the foreground
emission if the antenna is not aligned with the orientation of the
foreground polarization. On the other hand, a full-Stokes measurement
as proposed in this study can measure the intrinsically polarized
portion of the foreground and improves the accuracy of the foreground
spectrum measurement. Since the intrinsic polarization follows the
apparent foreground rotation relative to the antenna on the ground,
its contribution to the net projection-induced polarization can be
represented as a vector summation between the projection Stokes vector
$\bold{S}_{net}$ and the intrinsic Stokes vector $\bold{S}_{intr}$. As
a result, the total measured polarization $\bold{S}_{tot}$ also
contains a component of twice the sky revolution rate. A total-power
spectrum and a second harmonic spectrum can be reconstructed as
before. Assuming only the polarizations are linear,
\autoref{fig:intrinsic_fg_polarization_vector_illustration}
illustrates the vector summation between Stokes vectors can be
represented by a 2-D version of the Poincar\'e sphere with the $Q$ and
$U$ axes.

\subsection{Foreground spectral index variations and detection validation}
\label{sec:effect_spectral_index}
\begin{figure}
\centering 
\includegraphics[scale=.65]{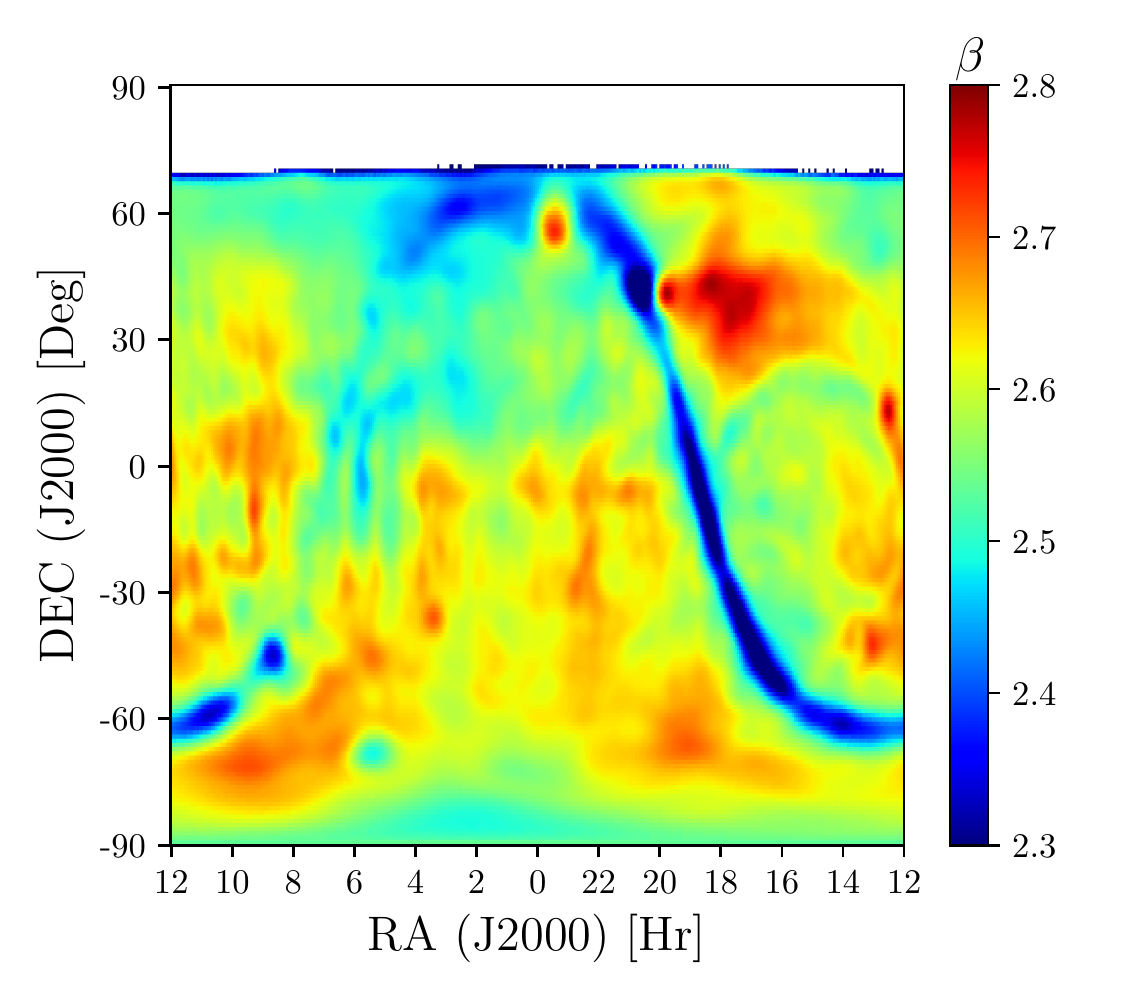} 
\caption{Foreground spectral index distribution is obtained from
  extrapolating between the Haslam all-sky survey at 408 MHz and a 45
  MHz map. The missing data around the NCP represents $\sim 4\%$ of
  the whole sky in the 45-MHz map.}\label{fig:fg_spectral_index_distr}
\end{figure}

The foreground continuum emission has been observed to have a
direction-dependent spectral index $\beta$ depending on the pointing,
i.e., $\beta = \beta(\theta,\phi)$. The spectral index distribution
can be estimated by extrapolating from a power law between the Haslam
all-sky survey at 408 MHz and a 45 MHz map \citep{guzman2011all} as
shown in \autoref{fig:fg_spectral_index_distr}. This may potentially
complicate the foreground measurement using the second-harmonic Stokes
spectrum.
By writing Equation \eqref{eq:pw_law_fg_spatial} with a non-constant
$\beta$ as
\begin{equation}
T_{fg}(\theta,\phi,\nu) = T_{\rm Haslam}(\theta,\phi)\left(\frac{\nu}{408\ {\rm
    MHz}}\right)^{-\beta(\theta,\phi)}
\label{eq:pw_law_fg_spatial}
\end{equation}
it is apparent the chromatic antenna beam pattern can couple to the
sky differently at different frequencies. As a result, the net
projection-induced polarization can vary and contribute some unwanted
spectral feature to the Stokes spectra, especially the second-harmonic
one, if the antenna beam is not corrected.

However, since the induced-polarization approach measures both the
total-power spectrum and the second harmonic spectrum simultaneously
from the same sky region, as the simulations suggested, the measured
mean spectral index of $S_{Q,2}^{\nu}$ is still identical to the
foreground spectral index in $S_{I,0}^{\nu}$ for any given mean value
of $\beta$ averaged over the sky. In fact, one of the biggest
challenges for the global signal experiment is to ensure such any
spectral structures detected in the residual are indeed from the
background signal. This degeneracy can be broken by carrying out
multiple observation at different sky regions. For example, on the
ground, if the residual spectral features measured at the NCP and SCP
are consistent, they are very likely originated from the same source,
and possibly the background itself.

\begin{table*}[htp]
  \caption{Summary of harmonic analysis for the projection-induced
    Stokes spectra.} \label{tab:harmonic_mode}
  \centering
  \begin{tabular}{ p{5cm} | c c c c c | p{5cm}} 
    \hline
    \hline  
    Signal Sources\T\B & \multicolumn{5}{c|}{Harmonic Order $n$} &  \multirow{2}{*}{Notes} \\
    \quad Stokes Spectrum: &  $S_{I}^{\nu}$ & $S_{I_{p}}^{\nu}$ & $S_{Q}^{\nu}$ & $S_{U}^{\nu}$ & $S_{V}^{\nu}$ & {}\\
    \hline
    21-cm background\T                       & 0   & \ldots  & \ldots & \ldots & \ldots  & for circular Gaussian beams\\
    Projection-induced polarization          & \ldots  & \ldots  & 2  & 2  & \ldots  & for circular Gaussian beams\\
    Beam gain distortion \& elongation & 0, 4 & 4 &H\footnote{H - possible
      high-order harmonics, i.e., $n>4$} &H& H & need correction with beam models\\

    Horizon obstruction                      & 0, 1    & 1   & 1   & 1  & 1  & latitude \& sky dependent \\
    Boresight pointing error                 & 4   &4   &H   &H  &H &\\
    RFI                                      & 0   &H   &H   &H   &H  & location \& time dependent\\
    Ionospheric distortion                   & 0   &H   &H   &H   &H  & time \& latitude dependent, pseudo-periodic\\
    \hline                                   
  \end{tabular}
\end{table*}

\subsection{RFI and ionospheric distortion}
\label{sec:other_distortion}
One of the major sources of contamination is RFI. Traditionally,
radio-quiet sites, such as South Africa's Karoo desert and Murchison
Radio-Astronomy Observatory in Western Australia, are chosen for
minimal contamination. Recently, DARE proposed to carry out
observations on the lunar farside when the spacecraft orbits around
the Moon \citep{burns2012probing}. Nonetheless, regardless how quiet a
site is, there is always low-level RFI which is either reflected or
scattered from satellites, space debris, and the ionosphere
\citep{offringa2013brightness}.

A common practice is to utilize RFI excision algorithm, such as one
based on kurtosis of the incoming signal's statistics
\citep{antoni2004spectral, de2009simplified, nita2010generalized}, to
remove the corrupted signal channels. This approach is more effective
to detect the strong instead of the low-level RFI, especially ones
that are equivalent to the system's noise level. Hence, unknown
spectral structures can also be introduced to the measured
spectrum. However, since most of the incoming RFI are
direction-dependent, they do not necessarily introduce any dynamic
variations to the measured linear Stokes parameters, unless they are
affected by the ionosphere during propagation. More importantly, the
RFI does not contain a $n=2$ component, hence does not have a major
effect on the second-harmonic spectrum used to constrain the
foreground spectrum.

The ionosphere has been well known for causing refraction, absorption,
and self-emission to radio propagation. Not until recently that
careful studies have started investigating the effects of ionospheric
distortion to the ground-based 21-cm measurements
\citep{vedantham2013chromatic, datta2016effects, sokolowski2015impact,
  rogers2015radiometric}. The ionosphere consists of plasma layers
with complex structures whose characteristics are position and time
dependent. The total electron content (TEC), which represents the
total electron column density for a given direction, varies throughout
the day. These variations are pseudo-periodic from one day to another,
but not repeatable and difficult to quantify. Current total-power
global 21-cm experiments \citep{sokolowski2015impact,
  rogers2015radiometric} have only provided limited understanding on
how these ionospheric effects can impact the accuracy and precision of
the global 21-cm signal. Attempts to utilize Global Positioning System
(GPS) ionospheric measurements are considered insufficient to
calibrate the ionospheric effect \citep{datta2016effects}. However, by
measuring the dynamic ionospheric variations, our polarimetric
approach may provide some constraints on their influence on the
overall sky measurement as a function of time and observed
frequency. A more thorough study in this area will be conducted in
future work.

\subsection{Summary}
\label{sec:summary}
We have shown that the harmonic decomposition of the dynamic Stokes
parameters provides additional information on different types of
measurement and systematic uncertainties. In most cases, the
experiment can be designed and configured to modulate the foreground
measurement in attempt to distinguish these perturbations from the
signal of interest. Based on our simulations and analysis,
\autoref{tab:harmonic_mode} provides a rudimentary list of harmonic
components that can potentially be used to identify different signal
sources and bound the uncertainty in constraining the foreground
spectrum. Harmonics at higher orders ($n>4$) are indicated as ``H''
for reference.

In analogy to the CMB anisotropy studies \citep[i.e.,
][]{tegmark1996method, ichiki2014cmb}, the foreground spatial
anisotropy and hence its induced-polarization can also be decomposed
into spherical harmonic $Y_l^m(\theta,\phi,\nu)$ with different modes
$l = 0,1,2,\ldots$ and $m = 0,\ldots,l$. We speculate that such
decomposition is complimentary to the harmonic analysis. In fact, by
combining the two, it may be possible to characterize different
contribution and systematics in the measurement by mapping out
different modes and harmonics, similar to
\autoref{tab:harmonic_mode}. This is outside the scope of the current
study.

\section{Conclusion}
\label{sec:conclusion}
A new observational approach is proposed to constrain the foreground
spectrum through measuring the projection-induced polarization of the
foreground anisotropy without assuming any sky model. Harmonic
analysis of the induced Stokes vector measured as a function of time
plays a key role in separating the dynamic foreground component from
the static background. The harmonic analysis also provides additional
information on instrumental systematics, and external contaminants
such as RFI and ionospheric distortions. This study established a
framework for developing our team's upcoming ground-based prototype,
the Cosmic Twilight Polarimeter (CTP), which will allow us to explore
some of the implementation aspects listed above to further evaluate
our approach.

By assuming an idealized instrument with a Gaussian beam, free from
RFI and ionospheric distortion, and the absorption feature of the
21-cm spectrum is present in the instrument passband, some of the key
aspects we found are:

\begin{itemize}
\item The spectral dependence of the foreground spectrum can be
  robustly measured in terms of the second-harmonic Stokes spectra
  $S_{Q,2}^{\nu}$ and $S_{U,2}^{\nu}$ retrieved from the
  frequency-dependent cyclic signature in the projection-induced
  polarization when the FOV is centered at a celestial pole.

\item The blind search for zero-crossing from the first derivative of
  the background 21-cm spectrum provides a unique initial condition in
  solving for the scaling factor between the Stokes spectra to be used
  for foreground subtraction, even without any knowledge of the
  background signal.

\item Chromatic distortion from the frequency-dependent antenna beam
  is one of the main instrumental effects that introduced unwanted
  spectral structures to confuse the foreground and the global 21-cm
  spectra. Careful beam calibration procedures, such as using beam models
  generated by sophisticated electromagnetic propagation simulation
  software and direct beam measurement, are required.

\item Environment factors and observation configuration, like ground
  effects and pointing error, contribute variations and confusions to
  the measurements in this polarimetric approach. Realistic
  assessment on instrument design and setup will help increasing the
  measurement sensitivity for the Stokes spectra.

\item As long as the foreground spectra from different sky regions,
  such as ones about the NCP and SCP, can be measured and subtracted
  independently, recovered global spectral structures will be
  consistent to a common homogeneous 21-cm background. This helps
  eliminating observational ambiguity and increases detection
  likelihood of the global 21-cm signal.

\end{itemize}

\section{Acknowledgments}
\label{sec:acknowledgments}
The National Radio Astronomy Observatory is a facility of the National
Science Foundation operated under cooperative agreement by Associated
Universities, Inc. Support for this work was provided by the NSF
through the Grote Reber Fellowship Program administered by Associated
Universities, Inc./National Radio Astronomy Observatory. This research
was also supported by the NASA Ames Research Center via Cooperative
Agreements NNA09DB30A, NNX15AD20A, and NNX16AF59G to J. O. Burns. The
authors like to especially thank J. Aguirre for his constructive
feedback on the formalism of our approach and discussions on other
aspects related to the foreground polarization. The authors would also
like to thank R. Monsalve for his comments on the instrumental aspects
of the approach. The authors gratefully acknowledge J. Mirocha for
providing the simulated global 21-cm data using ARES. B. D. Nhan is
also grateful to K. Tauscher and R. Monsalve for their assistance in
utilizing the sky map data.

\nocite{*} 
\bibliographystyle{aasjournal}
\bibliography{\myreferences}

\begin{thebibliography}{}
\expandafter\ifx\csname natexlab\endcsname\relax\def\natexlab#1{#1}\fi
\providecommand{\url}[1]{\href{#1}{#1}}

\bibitem[{Antoni(2004)}]{antoni2004spectral}
Antoni, J. 2004, 1167

\bibitem[{{Asad} {et~al.}(2015){Asad}, {Koopmans}, {Jeli{\'c}}, {Pandey},
  {Ghosh}, {Abdalla}, {Bernardi}, {Brentjens}, {de Bruyn}, {Bus}, {Ciardi},
  {Chapman}, {Daiboo}, {Fernandez}, {Harker}, {Iliev}, {Jensen},
  {Martinez-Rubi}, {Mellema}, {Mevius}, {Offringa}, {Patil}, {Schaye},
  {Thomas}, {van der Tol}, {Vedantham}, {Yatawatta}, \&
  {Zaroubi}}]{asad2015polarization}
{Asad}, K.~M.~B., {Koopmans}, L.~V.~E., {Jeli{\'c}}, V., {et~al.} 2015, \mnras,
  451, 3709

\bibitem[{Balanis(2016)}]{balanis2016antenna}
Balanis, C.~A. 2016, Antenna Theory: Analysis and Design (John Wiley \& Sons)

\bibitem[{Bernardi {et~al.}(2015)Bernardi, McQuinn, \&
  Greenhill}]{bernardi2015foreground}
Bernardi, G., McQuinn, M., \& Greenhill, L. 2015, The Astrophysical Journal,
  799, 90

\bibitem[{{Bittner} \& {Loeb}(2011)}]{bittner2011measuring}
{Bittner}, J.~M., \& {Loeb}, A. 2011, Journal of Cosmology and Astroparticle
  Physics, 4, 38

\bibitem[{Bowman \& Rogers(2010)}]{bowman2010lower}
Bowman, J.~D., \& Rogers, A.~E. 2010, Nature, 468, 796

\bibitem[{Bowman {et~al.}(2008)Bowman, Rogers, \& Hewitt}]{bowman2008toward}
Bowman, J.~D., Rogers, A.~E., \& Hewitt, J.~N. 2008, The Astrophysical Journal,
  676, 1

\bibitem[{Bowman {et~al.}(2013)Bowman, Cairns, Kaplan, Murphy, Oberoi,
  Staveley-Smith, Arcus, Barnes, Bernardi, Briggs,
  {et~al.}}]{bowman2013science}
Bowman, J.~D., Cairns, I., Kaplan, D.~L., {et~al.} 2013, Publications of the
  Astronomical Society of Australia, 30, e031

\bibitem[{{Burns} {et~al.}(2012){Burns}, {Lazio}, {Bale}, {Bowman}, {Bradley},
  {Carilli}, {Furlanetto}, {Harker}, {Loeb}, \& {Pritchard}}]{burns2012probing}
{Burns}, J.~O., {Lazio}, J., {Bale}, S., {et~al.} 2012, Advances in Space
  Research, 49, 433

\bibitem[{{Datta} {et~al.}(2016){Datta}, {Bradley}, {Burns}, {Harker},
  {Komjathy}, \& {Lazio}}]{datta2016effects}
{Datta}, A., {Bradley}, R., {Burns}, J.~O., {et~al.} 2016, \apj, 831, 6

\bibitem[{De~Roo(2009)}]{de2009simplified}
De~Roo, R.~D. 2009, IEEE Transactions on Geoscience and Remote Sensing, 47,
  3755

\bibitem[{Furlanetto(2006)}]{furlanetto2006global}
Furlanetto, S.~R. 2006, Monthly Notices of the Royal Astronomical Society, 371,
  867

\bibitem[{Furlanetto {et~al.}(2006)Furlanetto, Oh, \&
  Briggs}]{furlanetto2006cosmology}
Furlanetto, S.~R., Oh, S.~P., \& Briggs, F.~H. 2006, Physics Reports, 433, 181

\bibitem[{{Greenhill} \& {LEDA
  Collaboration}(2015)}]{greenhill2015constraining}
{Greenhill}, L.~J., \& {LEDA Collaboration}. 2015, in American Astronomical
  Society Meeting Abstracts, Vol. 225, American Astronomical Society Meeting
  Abstracts, 403.07

\bibitem[{Guzm{\'a}n {et~al.}(2011)Guzm{\'a}n, May, Alvarez, \&
  Maeda}]{guzman2011all}
Guzm{\'a}n, A.~E., May, J., Alvarez, H., \& Maeda, K. 2011, Astronomy \&
  Astrophysics, 525, A138

\bibitem[{Hamming(2012)}]{hamming2012numerical}
Hamming, R. 2012, Numerical Methods for Scientists and Engineers (Courier
  Corporation)

\bibitem[{Harker(2015)}]{harker2015selection}
Harker, G.~J. 2015, Monthly Notices of the Royal Astronomical Society: Letters,
  449, L21

\bibitem[{Harker {et~al.}(2012)Harker, Pritchard, Burns, \&
  Bowman}]{harker2012mcmc}
Harker, G.~J., Pritchard, J.~R., Burns, J.~O., \& Bowman, J.~D. 2012, Monthly
  Notices of the Royal Astronomical Society, 419, 1070

\bibitem[{Haslam {et~al.}(1982)Haslam, Salter, Stoffel, \&
  Wilson}]{haslam1982408}
Haslam, C., Salter, C., Stoffel, H., \& Wilson, W. 1982, Astronomy and
  Astrophysics Supplement Series, 47, 1

\bibitem[{{Heiles}(2002)}]{heiles2001heuristic}
{Heiles}, C. 2002, 278, 131

\bibitem[{Heinzel {et~al.}(2002)Heinzel, R{\"u}diger, \&
  Schilling}]{heinzel2002spectrum}
Heinzel, G., R{\"u}diger, A., \& Schilling, R. 2002

\bibitem[{Ichiki(2014)}]{ichiki2014cmb}
Ichiki, K. 2014, Progress of Theoretical and Experimental Physics, 2014, 06B109

\bibitem[{Jeli{\'c} {et~al.}(2014)Jeli{\'c}, De~Bruyn, Mevius, Abdalla, Asad,
  Bernardi, Brentjens, Bus, Chapman, Ciardi, {et~al.}}]{jelic2014initial}
Jeli{\'c}, V., De~Bruyn, A., Mevius, M., {et~al.} 2014, Astronomy \&
  Astrophysics, 568, A101

\bibitem[{Jeli{\'c} {et~al.}(2015)Jeli{\'c}, de~Bruyn, Pandey, Mevius,
  Haverkorn, Brentjens, Koopmans, Zaroubi, Abdalla, Asad,
  {et~al.}}]{jelic2015linear}
Jeli{\'c}, V., de~Bruyn, A., Pandey, V., {et~al.} 2015, Astronomy \&
  Astrophysics, 583, A137

\bibitem[{Johnson {et~al.}(2007)Johnson, Collins, Abroe, Ade, Bock, Borrill,
  Boscaleri, De~Bernardis, Hanany, Jaffe, {et~al.}}]{johnson2007maxipol}
Johnson, B., Collins, J., Abroe, M., {et~al.} 2007, The Astrophysical Journal,
  665, 42

\bibitem[{Kogut(2012)}]{kogut2012synchrotron}
Kogut, A. 2012, The Astrophysical Journal, 753, 110

\bibitem[{Kraus(1986)}]{kraus1986radio}
Kraus, J.~D. 1986, Radio Astronomy

\bibitem[{Kusaka {et~al.}(2014)Kusaka, Essinger-Hileman, Appel, Gallardo,
  Irwin, Jarosik, Nolta, Page, Parker, Raghunathan,
  {et~al.}}]{kusaka2014modulation}
Kusaka, A., Essinger-Hileman, T., Appel, J., {et~al.} 2014, Review of
  Scientific Instruments, 85, 024501

\bibitem[{Liu {et~al.}(2013)Liu, Pritchard, Tegmark, \& Loeb}]{liu2013global}
Liu, A., Pritchard, J.~R., Tegmark, M., \& Loeb, A. 2013, Physical Review D,
  87, 043002

\bibitem[{Mahesh {et~al.}(2014)Mahesh, Subrahmanyan, Shankar, \&
  Raghunathan}]{mahesh2014resistive}
Mahesh, N., Subrahmanyan, R., Shankar, N.~U., \& Raghunathan, A. 2014, arXiv
  preprint arXiv:1406.2585

\bibitem[{Mellema {et~al.}(2013)Mellema, Koopmans, Abdalla, Bernardi, Ciardi,
  Daiboo, de~Bruyn, Datta, Falcke, Ferrara, {et~al.}}]{mellema2013reionization}
Mellema, G., Koopmans, L.~V., Abdalla, F.~A., {et~al.} 2013, Experimental
  Astronomy, 36, 235

\bibitem[{Mirocha(2014)}]{mirocha2014decoding}
Mirocha, J. 2014, Monthly Notices of the Royal Astronomical Society, 443, 1211

\bibitem[{{Mirocha} {et~al.}(2017){Mirocha}, {Furlanetto}, \&
  {Sun}}]{mirocha2017global}
{Mirocha}, J., {Furlanetto}, S.~R., \& {Sun}, G. 2017, \mnras, 464, 1365

\bibitem[{Mirocha {et~al.}(2013)Mirocha, Harker, \&
  Burns}]{mirocha2013interpreting}
Mirocha, J., Harker, G.~J., \& Burns, J.~O. 2013, The Astrophysical Journal,
  777, 118

\bibitem[{Mirocha {et~al.}(2015)Mirocha, Harker, \&
  Burns}]{mirocha2015interpreting}
---. 2015, The Astrophysical Journal, 813, 11

\bibitem[{Mozdzen {et~al.}(2016)Mozdzen, Bowman, Monsalve, \&
  Rogers}]{mozdzen2016limits}
Mozdzen, T.~J., Bowman, J.~D., Monsalve, R.~A., \& Rogers, A.~E. 2016, Monthly
  Notices of the Royal Astronomical Society, 455, 3890

\bibitem[{Newell {et~al.}(2001)Newell, Greenwald, \&
  Ruohoniemi}]{newell2001role}
Newell, P.~T., Greenwald, R.~A., \& Ruohoniemi, J.~M. 2001, Reviews of
  Geophysics, 39, 137

\bibitem[{Nita \& Gary(2010)}]{nita2010generalized}
Nita, G.~M., \& Gary, D.~E. 2010, Monthly Notices of the Royal Astronomical
  Society: Letters, 406, L60

\bibitem[{Offringa {et~al.}(2013)Offringa, De~Bruyn, Zaroubi, Koopmans,
  Wijnholds, Abdalla, Brouw, Ciardi, Iliev, Harker,
  {et~al.}}]{offringa2013brightness}
Offringa, A., De~Bruyn, A., Zaroubi, S., {et~al.} 2013, Monthly Notices of the
  Royal Astronomical Society, stt1337

\bibitem[{Paciga {et~al.}(2013)Paciga, Albert, Bandura, Chang, Gupta, Hirata,
  Odegova, Pen, Peterson, Roy, {et~al.}}]{paciga2013simulation}
Paciga, G., Albert, J.~G., Bandura, K., {et~al.} 2013, Monthly Notices of the
  Royal Astronomical Society, stt753

\bibitem[{Parsons {et~al.}(2012)Parsons, Pober, Aguirre, Carilli, Jacobs, \&
  Moore}]{parsons2012per}
Parsons, A.~R., Pober, J.~C., Aguirre, J.~E., {et~al.} 2012, The Astrophysical
  Journal, 756, 165

\bibitem[{Parsons {et~al.}(2010)Parsons, Backer, Foster, Wright, Bradley,
  Gugliucci, Parashare, Benoit, Aguirre, Jacobs,
  {et~al.}}]{parsons2010precision}
Parsons, A.~R., Backer, D.~C., Foster, G.~S., {et~al.} 2010, The Astronomical
  Journal, 139, 1468

\bibitem[{Patra {et~al.}(2013)Patra, Subrahmanyan, Raghunathan, \&
  Shankar}]{patra2013saras}
Patra, N., Subrahmanyan, R., Raghunathan, A., \& Shankar, N.~U. 2013,
  Experimental Astronomy, 36, 319

\bibitem[{Petrovic \& Oh(2011)}]{petrovic2011systematic}
Petrovic, N., \& Oh, S.~P. 2011, Monthly Notices of the Royal Astronomical
  Society, 413, 2103

\bibitem[{{Pober} {et~al.}(2014){Pober}, {Liu}, {Dillon}, {Aguirre}, {Bowman},
  {Bradley}, {Carilli}, {DeBoer}, {Hewitt}, {Jacobs}, {McQuinn}, {Morales},
  {Parsons}, {Tegmark}, \& {Werthimer}}]{pober2014what}
{Pober}, J.~C., {Liu}, A., {Dillon}, J.~S., {et~al.} 2014, \apj, 782, 66

\bibitem[{{Pritchard} \& {Loeb}(2010)}]{pritchard2010constraining}
{Pritchard}, J.~R., \& {Loeb}, A. 2010, \prd, 82, 023006

\bibitem[{Pritchard \& Loeb(2012)}]{pritchard201221}
Pritchard, J.~R., \& Loeb, A. 2012, Reports on Progress in Physics, 75, 086901

\bibitem[{Rogers {et~al.}(2015)Rogers, Bowman, Vierinen, Monsalve, \&
  Mozdzen}]{rogers2015radiometric}
Rogers, A.~E., Bowman, J.~D., Vierinen, J., Monsalve, R., \& Mozdzen, T. 2015,
  Radio Science, 50, 130

\bibitem[{{Shaver} {et~al.}(1999){Shaver}, {Windhorst}, {Madau}, \& {de
  Bruyn}}]{shaver1999can}
{Shaver}, P.~A., {Windhorst}, R.~A., {Madau}, P., \& {de Bruyn}, A.~G. 1999,
  \aap, 345, 380

\bibitem[{Sokolowski {et~al.}(2015{\natexlab{a}})Sokolowski, Tremblay, Wayth,
  Tingay, Clarke, Roberts, Waterson, Ekers, Hall, Lewis,
  {et~al.}}]{sokolowski2015bighorns}
Sokolowski, M., Tremblay, S.~E., Wayth, R.~B., {et~al.} 2015{\natexlab{a}},
  Publications of the Astronomical Society of Australia, 32, e004

\bibitem[{Sokolowski {et~al.}(2015{\natexlab{b}})Sokolowski, Wayth, Tremblay,
  Tingay, Waterson, Tickner, Emrich, Schlagenhaufer, Kenney, \&
  Padhi}]{sokolowski2015impact}
Sokolowski, M., Wayth, R.~B., Tremblay, S.~E., {et~al.} 2015{\natexlab{b}}, The
  Astrophysical Journal, 813, 18

\bibitem[{{Switzer} \& {Liu}(2014)}]{switzer2014erasing}
{Switzer}, E.~R., \& {Liu}, A. 2014, \apj, 793, 102

\bibitem[{{Tegmark} \& {Efstathiou}(1996)}]{tegmark1996method}
{Tegmark}, M., \& {Efstathiou}, G. 1996, \mnras, 281, 1297

\bibitem[{Tegmark {et~al.}(2000)Tegmark, Eisenstein, Hu, \&
  de~Oliveira-Costa}]{tegmark2000foregrounds}
Tegmark, M., Eisenstein, D.~J., Hu, W., \& de~Oliveira-Costa, A. 2000, The
  Astrophysical Journal, 530, 133

\bibitem[{Tingay {et~al.}(2013)Tingay, Goeke, Bowman, Emrich, Ord, Mitchell,
  Morales, Booler, Crosse, Wayth, {et~al.}}]{tingay2013murchison}
Tingay, S., Goeke, R., Bowman, J.~D., {et~al.} 2013, Publications of the
  Astronomical Society of Australia, 30, e007

\bibitem[{Trippe(2014)}]{trippe2014polarization}
Trippe, S. 2014, Journal of the Korean Astronomical Society, 47, 15

\bibitem[{Van~Haarlem {et~al.}(2013)Van~Haarlem, Wise, Gunst, Heald, McKean,
  Hessels, De~Bruyn, Nijboer, Swinbank, Fallows, {et~al.}}]{van2013lofar}
Van~Haarlem, M., Wise, M., Gunst, A., {et~al.} 2013, Astronomy \& astrophysics,
  556, A2

\bibitem[{Vedantham {et~al.}(2013)Vedantham, Koopmans, de~Bruyn, Wijnholds,
  Ciardi, \& Brentjens}]{vedantham2013chromatic}
Vedantham, H., Koopmans, L., de~Bruyn, A., {et~al.} 2013, Monthly Notices of
  the Royal Astronomical Society, stt1878

\bibitem[{Voytek {et~al.}(2014)Voytek, Natarajan, Garc{\'\i}a, Peterson, \&
  L{\'o}pez-Cruz}]{voytek2014probing}
Voytek, T.~C., Natarajan, A., Garc{\'\i}a, J. M.~J., Peterson, J.~B., \&
  L{\'o}pez-Cruz, O. 2014, The Astrophysical Journal Letters, 782, L9

\bibitem[{Wilson {et~al.}(2009)Wilson, Rohlfs, \&
  H{\"u}ttemeister}]{wilson2009tools}
Wilson, T.~L., Rohlfs, K., \& H{\"u}ttemeister, S. 2009, Tools of Radio
  Astronomy, Vol.~86 (Springer)

\bibitem[{Wolleben {et~al.}(2006)Wolleben, Landecker, Reich, \&
  Wielebinski}]{wolleben2006absolutely}
Wolleben, M., Landecker, T., Reich, W., \& Wielebinski, R. 2006, Astronomy \&
  Astrophysics, 448, 411

\end{thebibliography}
\end{document}